\documentclass[draftcls,onecolumn,12pt]{IEEEtran}
\usepackage{graphicx}

\usepackage{amsmath,amsthm}

\usepackage{amsfonts}

\usepackage{cases}
\usepackage{subeqnarray}

\usepackage{url}
\usepackage{subfigure}

\usepackage{booktabs}

\usepackage{setspace}

\ifCLASSINFOpdf
\else
\fi
\begin{document}
%
\title{\LARGE A Cross-Layer Perspective on Energy Harvesting Aided Green Communications over Fading Channels}
%
%
%
%

\author{Tian~Zhang,
Wei~Chen,~\IEEEmembership{Senior Member,~IEEE,}
        Zhu Han,~\IEEEmembership{Senior Member,~IEEE,}
        and~Zhigang~Cao,~\IEEEmembership{Senior Member,~IEEE}
\thanks{The material in this paper was presented in part at the {\em IEEE INFOCOM Workshop CCSES'13}, Turin, Italy, Apr. 2013 \cite{INFOCOMW13: Tian
Zhang W. Chen Z. Han and Z. Cao}.}
\thanks{ T. Zhang is with the School of Information Science and Engineering, Shandong University, Jinan 250100, China. He is also with the Department
of Electronic Engineering, Tsinghua University.
E-mail: tianzhang.ee@gmail.com}
\thanks{ W. Chen and Z. Cao are with the Department of Electronic Engineering, Tsinghua University, Beijing 100084, China.
E-mail: \{wchen, czg-dee\}@tsinghua.edu.cn.}
\thanks{ Z. Han is with the Department of Electrical and Computer Engineering, University
of Houston, Houston, TX 77004, USA.
E-mail: zhan2@uh.edu}
}

\begin{spacing}{1}
\maketitle
\vspace{-1em}
\thispagestyle{empty}
\begin{spacing}{1.3}
\begin{abstract}
In this paper, we consider the power allocation of the physical layer and the buffer delay of the upper application layer in energy harvesting green
networks. The total power required for reliable transmission includes the transmission power and the circuit power. The harvested power (which is
stored in a battery) and the grid power constitute the power resource. The uncertainty of data generated from the upper layer, the intermittence of
the harvested energy, and the variation of the fading channel are taken into account and described as independent Markov processes. In each
transmission, the transmitter decides the transmission rate as well as the allocated power from the battery, and the rest of the required power will
be supplied by the power grid. The objective is to find an allocation sequence of transmission rate and battery power to minimize the long-term
average buffer delay under the average grid power constraint. A stochastic optimization problem is formulated accordingly to find such transmission
rate and battery power sequence. Furthermore, the optimization problem is reformulated as a constrained Markov decision process (MDP) problem whose
policy is a two-dimensional vector with the transmission rate and the power allocation of the battery as its elements. We prove that the optimal
policy of the constrained MDP can be obtained by solving the unconstrained MDP. Then we focus on the analysis of the unconstrained average-cost MDP.
The structural properties of the average optimal policy are derived.
Moreover, we discuss the relations between elements of the two-dimensional policy. Next, based on the theoretical analysis, the algorithm to find the
constrained optimal policy is presented for the finite state space scenario. In addition, heuristic policies (two deterministic policies and a mixed policy) with low-complexity are given for the general state space.
Finally, simulations are performed under these policies to demonstrate the effectiveness.

\end{abstract}

\end{spacing}
\end{spacing}

\IEEEdisplaynotcompsoctitleabstractindextext

%
\IEEEpeerreviewmaketitle
\newpage
\setcounter{page}{1}

\section{Introduction}
%
%

%
%
%
%
\IEEEPARstart{R}{apid} wireless communication industry development has led to a dramatic increase of energy consumption in wireless networks, and such
an increasing energy consumption produces a series of energetic and environmental problems.
Recently, green communications, which aims at enhancing energy efficiency and carbon emission reduction, have received considerable
attention \cite{WCMC09:G. Miao N. Himayat Y. (Geoffrey) Li and A. Swami}-\cite{WC11:G. Auer V. Giannini C. Desset}. In the energy-efficient design for
wireless communications, the total energy consumption includes not only the transmission
energy but also the circuit energy consumption \cite{TWC05:S. Cui A. Goldsmith A. Bahai}.
\par
As a preferred choice supporting green
communications, energy harvesting techniques such as photovoltaic solar cells become popular for the ability to prolong the lifetime of the battery
and the lifetime of wireless networks thereby. There have been a lot of researches in wireless networks with energy harvesting nodes.
In \cite{WC07:D. Niyato E. Hossain M. Rashid and V. Bhargava}, an optimal energy
management policy for a solar-powered sensor node was proposed. The policy uses a sleep and wakeup strategy for energy conservation. In \cite{TWC10:V.
Sharma U. Mukherji V. Joseph and S. Gupta}, throughput optimal and mean delay optimal energy management policies were studied for a single energy
harvesting sensor node. The Shannon capacity of an energy harvesting
sensor node transmitting over an AWGN channel was obtained in \cite{ISIT11: R. Rajesh V. Sharm and P. Viswanath}. In \cite{ITA12:N. Michelusi K.
Stamatiou and M. Zorzi}, the optimal binary transmission policies were studied under i.i.d. Bernoulli energy arrivals.
In \cite{TC12:Z. Wang A. Tajer and X. Wang}, the long-term average communication
reliability optimization problem was studied for the system
of energy-harvesting active networked tags (EnHANTs). In \cite{TSP12:C. K. Ho and R. Zhang} and \cite{forJSAC12:J. Xu and R. Zhang},
throughput-maximal schemes of energy allocation
for wireless communications with energy harvesting constraints are studied.
\par
Resource allocation is a fundamental problem in wireless communications \cite{BOOK08:Z. Han and K. J. R. Liu}. Generally, resource consumption
reduction and quality of service (QoS) improvement are two conflicting objectives in a resource allocation problem. There has been some interests in
analyzing the power allocation and delay performance from the cross-layer perspective.
In \cite{thesis00: R. Berry} and \cite{TIT02: R. Berry and R. Gallager}, the tradeoff between the average required power for reliable transmission at
the physical layer and the mean delay at the network layer was studied in fading channels. The adaptive control policies utilize information on both
queue state and channel state, and some structural results for the
optimal policy were derived.
In \cite{TIT08: M. Goyal A. Kumar and V. Sharma}, the authors derived the improved results upon
these obtained in \cite{TIT02: R. Berry and R. Gallager}. They considered the optimization problem aiming to minimize the delay in
the transmitter buffer under an average transmitter power constraint. The existence of stationary average
optimal policy was proved and  some structural results were obtained. In \cite{OR05: B. Ata}, the fading channel was simplified to a static channel,
and the explicit optimal control policy was characterized.
\par
In \cite{TIT02: R. Berry and R. Gallager}-\cite{OR05: B. Ata}, only the transmission power is considered. However, as shown in \cite{WCMC09:G. Miao N.
Himayat Y. (Geoffrey) Li and A. Swami}, the transmission strategy changes when taking the circuit power into account. Then a natural problem is {\em
what about the power and delay when considering both transmission power and circuit power.} Meanwhile, as energy allocation of the battery plays a
central role in the transmission strategy of energy harvesting nodes, {\em how the energy allocation strategy of the battery will affect the power and
delay?}
\par
In this paper, we consider the power allocation in the physical layer and the delay performance in the upper application layer in green wireless
networks with energy harvesting nodes. The data are generated in the application layer, and placed in a buffer at the transmitter. The transmitter
periodically removes some data from the buffer, and transmits the data to the receiver. The required power for reliable transmission takes both
transmission power and circuit power into account, and the power resource makes up of the harvested power and grid power.
The harvested energy arrives randomly, and there is a constraint on the average grid power. The objective is to minimize the average delay in the
buffer with a constrained average grid power and random battery energy.
Since the required power for each transmission can be supplied from both the battery and the grid, the policy is two-dimensional, i.e., the rate as
well as the allocation of the battery energy (the grid power allocation is then the total required power minus the allocated battery power), in the
formulated optimization problem.
\par
Specifically, the main contributions of the paper can be summarized as follows.
\begin{itemize}[]
\item We consider the delay-optimal power allocation in the framework of green communications over fading channels, where the power comes from
    both power grid and harvesting devices. The data arrival process, the harvested energy arrival process, and the channel process are
    Markovian.
    A stochastic optimization problem is formulated to find a transmission rate and battery power allocation sequence to minimize the long-run
    average buffer delay under the constraint on the average grid power.
\item
    We reformulate the optimization problem as a constrained Markov decision process (MDP) problem, in which the state and action are defined.
    The state includes the queue state, the battery state (i.e., the stored energy in the battery), the channel state, the data arrival, and the
    harvested energy arrival. The action consists of the transmission rate and the power allocation from the battery.
    Using the Lagrangian methodology, the constrained MDP can be relaxed to an unconstrained problem (UP), which is an average cost MDP. We prove
    that the optimal solution of the constrained MDP can be derived by solving the UP with one or two Lagrangian multipliers. Then we focus on
    the optimal policy of the average cost MDP (i.e., UP). We verify the existence of the optimal stationary policy of the average cost MDP and it
    can be obtained from the corresponding  discount cost MDP. We derive two necessary conditions for the optimal policy of the average cost MDP
    (average cost optimal policy). Under certain conditions, the policy that serving nothing and allocating no energy from the battery
    is an average cost optimal policy. We also prove that serving everything combined with allocating the minimal of the total required power and
    total energy in the battery are an average cost optimal policy under other certain conditions. The monotonicities of the optimal object value with respect to Lagrangian multiplier and optimal policy regarding the state are investigated, respectively.
\item We analyze the relations between the transmission rate and the power allocation from the battery. We find that given the transmission rate
    policy, the optimal battery power allocation policy is the greedy policy in some scenario. For general scenario, we propose a sufficient
    condition under which the optimal policy of two-dimensional MDP problem can be decomposed to the optimal policy of an MDP problem with the
    policy to be the transmission rate only in addition with the greedy battery power allocation policy.
\item On the basis of the theoretical investigation, we propose an algorithm to find the constrained optimal policy under the finite state case.
In addition, we propose three heuristic policies for the constrained MDP with the general state case: radical policy, conservative policy, and mixed
policy.
\end{itemize}
\par
The remainder of the paper is organized as follows. In Section II, the system model is described, and a mean buffer delay minimization problem with
average grid power constraint is formulated. In Section III, the optimization problem
is re-formulated as a constrained MDP and the optimal two-dimensional policy of the constrained MDP is investigated. Next,
we discuss the relations between elements of the two-dimensional policy in Section \ref{dimesion reduction}. Based on the theoretical analysis, the algorithm to find the constrained optimal policy under the finite state space and heuristic policies for the general state space are proposed in Section \ref{Proposedpolicies}. Simulations are performed in Section VI. Finally, Section VII
concludes the paper.


\section{System model and problem formulation}

\begin{figure}[!t]
\centering
\includegraphics[width=3.0in]{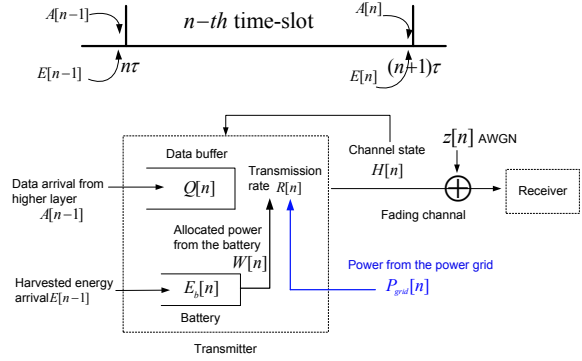}
\vspace{-1em}
\caption{System model}
\vspace{-1em}
\label{fig_systemmodel}
\end{figure}

We consider a slotted-time model of a point-to-point block fading channel. The length of a time-slot is $\tau$ units. The $n$-th time-slot is the time
interval $\big[n\tau,(n+1)\tau \big)$. The channel gain remains static in each slot, and changes between different slots. The sequence of the channel
gains is a finite-state ergodic Markov chain $\{H[n]\}$. The transmitter is assumed to have perfect channel state information (CSI). As shown in Fig.
\ref{fig_systemmodel}, at the end of the $n$-th slot, the higher layer generates $A[n]$ packets and they are stored in a buffer before transmission.
It is assumed that each packet is with $b$ bits and $\{A[n]\}$ is a finite-state ergodic Markov chain.
We assume that the transmitter is equipped with an energy harvesting device and it can also get power from the power grid.\footnote{Grid power with average constraint is to guarantee user's QoS (delay). Specifically, due to the causality of harvested energy, the transmitter should accumulate a sufficient amount of energy before each packet transmission. Then the waiting time could be undesirably long since the randomness of harvested energy arrival. In contrast, when the grid power is available, even if the battery energy is insufficient, the transmitter could use the grid power to transmit packet. Hence, the user's QoS can be guaranteed.} The harvested energy
arrives at each end of the slot according to a finite-state ergodic Markov chain $\{E[n]\}$, and the harvested energy will be stored in a battery
before consumption. There exists a long run average constraint on the grid power at the transmitter.
At the beginning of the $n$-th time slot, the transmitter chooses $R[n]$ packets from the buffer and transmits to the receiver.\footnote{$R[n]$ is the transmission
rate of the $n$-th timeslot with unit packets/timeslot.} We assume the additive white Gaussian noise (AWGN) at the receiver is with zero mean and
variance $\sigma^2$.
In green communications, the total power required for reliable transmissions\footnote{In the paper, \lq\lq reliable transmission\rq\rq~ means totally
error-free according to capacity arguments.} of $R[n]$ packets in the $n$-th time-slot is \cite{WCMC09:G. Miao N. Himayat Y. (Geoffrey) Li and A.
Swami}
 \begin{eqnarray}\label{relation between required power and rate}
 P(X[n],R[n])=\rho \frac{\sigma^2}{H[n]}(e^{\theta R[n]}-1)+\Delta(R[n]),
 \end{eqnarray}
where $X[n]$ is the system state that will be defined later, $\rho \ge 1$ is a constant, $\theta=\frac{2\ln(2)b}{N}$ with $N$ being the channel uses
in each time-slot, and
 \begin{eqnarray}
 \Delta (R[n]) = \left\{ \begin{array}{l}
 C_{} ,R[n] \ne 0 ;\\
 0_{} ,R[n] = 0 ,\\
 \end{array} \right.
  \end{eqnarray}
where $C \ge 0$ is a constant. In particular, $\rho =1$ and $C=0$ when no circuit power is taken into account.
In the transmission during the $n$-th timeslot, the transmitter allocates $W[n]$ power from the battery, and the remaining power will be supplied by the
power grid.
Denote $Q[n]$ as the queue length of the buffer at instance $n\tau$,
 the evolution equation for the buffer length is
 \begin{eqnarray}\label{evolution equation for the buffer length}
 Q[n+1]=Q[n]-R[n]+A[n].
\end{eqnarray}
Assume that the capacity for the battery is $E_{max}$.
Denote the battery's stored energy at instance $n\tau$ as $E_b[n]$, then the evolution equation for harvested energy in the battery can be given by
 \begin{eqnarray}\label{energy evolution in the battery}
 E_{b}[n+1]&=&\min\left\{E_{b}[n]-W[n]\tau+E[n],E_{max}\right\}:=(E_{b}[n]-W[n]\tau+E[n])^{-}.
 \end{eqnarray}
 The objective is to find a rate and battery power allocation sequence that minimizes the mean buffer delay under the constraint on the long-run
 average grid power $\bar{\mathcal{P}}$, and
 the stochastic optimization problem is given by
\begin{eqnarray} \label{original optimization problem}
\min_{\left\{(R[n],W[n])\right\}_{n=1}^\infty} \mathop {\lim \sup }\limits_{n \to \infty} \frac{1}{n}\mathbb{E}\left[\sum\limits_{k =0}^{n-1}
{Q[k]}\right]
\end{eqnarray}
\begin{subequations}
\begin{numcases}{\mbox{s.t.}}
\mathop {\lim \sup }\limits_{n\to \infty} \frac{1}{n}\mathbb{E}\left[\sum\limits_{k =0}^{n-1} P_{grid}[k]\right] \le \bar{\mathcal{P}},\label{original
avarge grid Cons}\\
R[k]\le Q[k],\label{original Cons1}\\
W[k]\tau \le E_{b}[k],\label{original Cons2}
\end{numcases}
\end{subequations}
 where $P_{grid}[k]$ is the power from the power grid,
\begin{eqnarray}\label{Required power is from grid and harvesting}
P(X[k],R[k])=P_{grid}[k]+W[k].
\end{eqnarray}
\theoremstyle{definition} \newtheorem{theorem}{Theorem}
\theoremstyle{definition} \newtheorem{remark}{Remark}
\theoremstyle{definition} \newtheorem{lemma}{Lemma}
\theoremstyle{definition} \newtheorem{property}{Property}
\theoremstyle{definition} \newtheorem{proposition}{Proposition}
\theoremstyle{definition} \newtheorem{claim}{Claim}
\theoremstyle{definition} \newtheorem{conjecture}{Conjecture}

\section{Analysis of the formulated stochastic optimization problem}\label{Analysis of the formulated stochastic optimization problem}
\begin{figure}[!t]
\centering
\includegraphics[width=4.0in,angle=90]{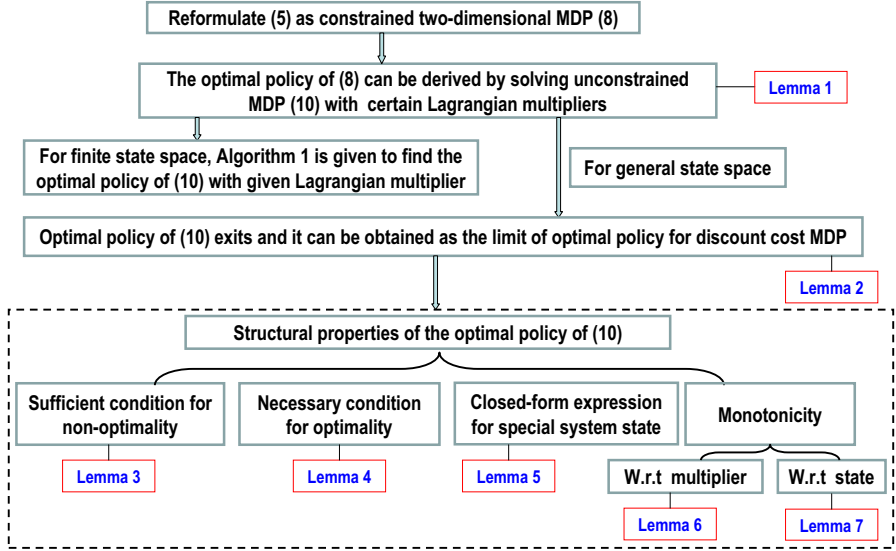}
\caption{Analysis structure of Section \ref{Analysis of the formulated stochastic optimization problem}}
\vspace{-1em}
\label{Analysis structure when the battery capacity is finite}
\vspace{-1em}
\end{figure}

In this section, we first reconstruct the problem (\ref{original optimization problem}) as a constrained two-dimensional (i.e., rate and battery power
allocation) MDP. Second, we prove that the constrained two-dimensional MDP can be transformed to unconstrained MDP by the Lagrangian method
in Section \ref{original optimal}. Then we focus on the analysis of the unconstrained MDP in Section \ref{AnalyofUP}. We verify the existence of
the stationary policy for the unconstrained MDP (which is an average cost MDP) in Section \ref{existence of optimal}. Next, we investigate the optimal
policy of the average cost MDP, and structural properties of the average cost optimal policy are derived in Section \ref{average optimal}. For better
readability, the analysis flowchart for this section is illustrated in Fig. \ref{Analysis structure when the battery capacity is finite}.

\subsection{Reconstruction as a constrained two-dimensional MDP}\label{reformulate as cMDP}
Define the state as
$X[n]:=\left(Q[n],H[n],A[n],E_{b}[n],E[n]\right)$ with state space $\mathcal{X}$
and the action as $\mathfrak{A}[n]:=\left(R[n],W[n]\right)$ with action space $\mathcal{A}$, respectively.\footnote{The system state includes the
buffer queue length, channel gain, data arrival, energy in the battery, and harvested energy arrival. The action includes the allocated rate and the
allocated battery energy.} Then $\left\{X[n], \mathfrak{A}[n]\right\}$ can be viewed as a Markov decision process (MDP).
The feasible action $(r,w)$ in a state $x=(q,h,a,e_b,e) \in \mathcal{X}$ belongs to $\mathcal{A}(x)=\{0,1,\cdots,q\}\times
\{0,\frac{1}{\tau},\cdots,\frac{e_b}{\tau}\}$.\footnote{The harvested energy has been discretized.}
Define a policy $\pi=(\pi_0,\pi_1,\cdots)$ that $\pi_n$ generates an action $\left(r[n],w[n]\right)$ with a probability at instant $n\tau$
\cite{Book99:E. Altman}\cite{Book02:E. Feinberg and A. Shwartz}. We denote the set of all policies as $\Pi$.
Specially, a stationary deterministic policy is $\pi=(g,g,\cdots)$, where $g$ is a measurable mapping from $\mathcal{X}$ to $\mathcal{A}$ such that
$g(x) \in \mathcal{A}(x)$ for all $x \in \mathcal{X}$.
Then, (\ref{original optimization problem}) is reformulated as the constrained MDP to find the two-dimensional (i.e., rate and battery power
allocation) optimal policy.
\begin{eqnarray} \label{original MDP}
\min_{\pi \in \Pi}B_x^{\pi} =\mathop {\lim \sup }\limits_{n\to \infty} \frac{1}{n}\mathbb{E}_{x}^{\pi}\left[\sum\limits_{k =0}^{n-1} {Q[k]}\right]
\end{eqnarray}
\begin{equation}
\mbox{s.t.}\quad K_x^{\pi} = \mathop {\lim \sup }\limits_{n\to \infty} \frac{1}{n}\mathbb{E}_{x}^{\pi}\left[\sum\limits_{k =0}^{n-1}
P_{grid}[k]\right] \le \bar{\mathcal{P}},\\
\end{equation}
%
%
where the subscript $x=(q,h,a,e_b,e) \in \mathcal{X}$ is the initial system state.

\subsection{Transformation to unconstrained MDP}\label{original optimal}
Define $P_{grid}(x,r,w):=\max\{P(x,r)-w,0\}:=(P(x,r)-w)^+$ and $f_{\beta}(x,r,w):=q+\beta P_{grid}(x,r,w)$ with $\beta>0$.
Then we have a family of the following unconstrained problem (UP$_\beta$).
 \begin{eqnarray}\label{UP_beta}
 \min_{\pi}  J_{x}^{\pi}(\beta):=\mathop {\lim \sup }\limits_{n\to \infty} \frac{1}{n}\mathbb{E}_x^\pi  \left[\sum\limits_{k = 0}^{n - 1} {f_\beta
 (X[k],R[k],W[k])}\right].
 \end{eqnarray}
In UP$_\beta$, $f_\beta (X[k],R[k],W[k])$ is the one-step cost in the $k$-th time-slot.
\par
\emph{Remark: UP$_\beta$ is an average cost MDP. Its optimal solution is called the average cost optimal policy.}
\par
The following lemma gives the relation between UP$_\beta$ and the constrained two-dimensional MDP (\ref{original MDP}).
\begin{lemma}\label{OP of the general}
When there exists a $\beta_0 >0$ that the optimal policy of UP$_{\beta_0}$ has an average grid power consumption equal to $\bar{\mathcal{P}}$,
the optimal solution of  UP$_\beta$ is optimal for the constrained MDP in (\ref{original MDP}). Otherwise, there exit a $\beta^+>0$ and a $\beta^->0$.
The optimal policy for the constrained MDP (\ref{original MDP}) is as follows: at each decision epoch, choose $\pi^-$ with a certain probability $q$
and $\pi^+$ with probability $1-q$, where $\pi^+$ and $\pi^-$ are the optimal policies obtained for UP$_{\beta^+}$ and UP$_{\beta^-}$, respectively.
$q$ depends on $\bar{\mathcal{P}}$ and the grid power consumptions of the two policies. \end{lemma}
\begin{IEEEproof}
See Appendix \ref{Proof of Lemma OP of the general}.
\end{IEEEproof}
Lemma \ref{OP of the general} reveals that the solution of (\ref{original MDP}) can be obtained by solving UP$_\beta$ with one or two $\beta$.
In the following, we focus on the analysis of the unconstrained MDP, UP$_\beta$.
\subsection{Analysis of the unconstrained MDP}\label{AnalyofUP}

\subsubsection{Existence of the optimal policy}\label{existence of optimal}
 Define a discount cost MDP with discount factor $\alpha \in (0,1)$ corresponding to UP$_\beta$ for each initial state $x=(q,h,a,e_b,e)$, with value
 function
  \begin{eqnarray}
  V_\alpha(x)=
  \min_{\pi}  \mathbb{E}_x^\pi \left[\sum_{k=0}^{\infty} \alpha^{k}\left(Q[k]+\beta P_{grid}(X[k],R[k],W[k])\right)\right].
  \end{eqnarray}
The optimal solution for the discounted problem is referred to as a discount optimal policy.
 \par
 The following lemma reveals the existence of the stationary policy. Furthermore, it derives how to obtain the optimal solution.
\begin{lemma}\label{discout to UP}
There exists a stationary deterministic policy that solves UP$_{\beta}$ with a $\beta>0$, and it can be obtained as a limit of discount optimal
policies as the discount factor increases to one.
\end{lemma}
\begin{IEEEproof}
See Appendix \ref{Proof of Lemma discout to UP}.
\end{IEEEproof}

Following the proof of Lema \ref{discout to UP}, we can also derive that the optimal $J_{x}^{\pi^*}(\beta)$ is independent of the initial state
$x$. Thus we can rewrite $J_{x}^{\pi^*}(\beta)$ as $J^{\pi^*}(\beta)$.

If the state is finite (Specifically, the data buffer state is finite), the relative value iteration algorithm (Algorithm 1) \cite{Book94:M. L. Puterman} can be utilized to find the optimal policy of the unconstrained MDP
UP$_{\beta}$ with given $\beta$. However, we are interested in deriving structural results on the optimal policies under general state
space\footnote{The number of data buffer states can be infinite, then the state number can be infinite in the paper.} and not simply solving the unconstrained problem with finite state space.
Furthermore, some structural results are useful to solve the constrained MDP (Section V).\par
    \begin{table}[]
    \caption{}\label{Value iteration algorithm of finding the optimal policy for UPbeta}
    \centering
    \begin{tabular}{lcl}
     \toprule
     \textbf{Algorithm 1: Relative value iteration algorithm of finding the optimal policy for UP$_\beta$} \\
     \midrule
    Step 1:  Select initial value $V^0$, choose reference state $x^*\in \mathcal{X}$, specify $\epsilon$, and set $n=0$\\
    Step 2: For each $x=(q,h,a,e_b,e) \in \mathcal{X}$, compute $V^{n+1}(x,\beta)$ by  \\
    $V^{n+1}(x,\beta)=\min\limits_{(r,w)\in
    \mathcal{A}(x)}\Big\{f_{\beta}(x,r,w)+\sum\limits_{x^{'}=(q^{'},h^{'},a^{'},e_b^{'},e^{'})\in\mathcal{X}}p(x^{'}|x,(r,w))V^{n}(x^{'},\beta)\Big\}$\\
    where
    $p(x^{'}|x,(r,w))=\delta(q-r+a-q')\delta(e_b-w+e-e_b^{'})p(h^{'}|h)p(a^{'}|a)p(e^{'}|e)$ \\
    is the transition probability, $\delta(0)=1$ and $\delta(x)=0$ when $x\ne 0$.  \\
    Step 3: Normalize $V^{n+1}(x,\beta)$ for each $x\in\mathcal{X}$ as
    $V^{n+1}(x,\beta)=V^{n+1}(x,\beta)-V^{n+1}(x^*,\beta)$  \\
    Step 4: If $|V^{n+1}-V^n|<\epsilon$, go to next Step. Otherwise, $n=n+1$ and go to Step 2.  \\
    Step 5: For each $x\in \mathcal{X}$, choose the policy according to   \\
     $\pi(x,\beta)=  \mathrm{arg}\min\limits_{(r,w)\in
     \mathcal{A}(x)}\Big\{f_{\beta}(x,r,w)+\sum\limits_{x^{'}\in\mathcal{X}}p(x^{'}|x,(r,w))V^{n}(x^{'},\beta)\Big\}$
                          \\
     \bottomrule
    \end{tabular}
    \end{table}

\subsubsection{Structural properties}\label{average optimal}

%

The average optimal policy are discussed in the subsection. First, the sufficient condition for non-optimality, necessary condition for
optimality, and the closed-form expressions of optimal policy in special system states are given.
\begin{lemma}\label{Sufficient condition for the non-optimality}
 In state $x=(q,h,a,e_b,e)$, $(r(x),w(x))$ is not the average cost optimal policy if $q-r(x)\ne 0$ and $e_b-w(x)+e>E_{max}$.
\end{lemma}
\begin{IEEEproof}
When a policy results in battery overflow (i.e., $e_b-w(x)+e>E_{max}$) and non-emptiness of the buffer (i.e., $q-r(x)\ne 0$), then in terms of the
average cost performance, the policy can be improved  by using the overflowed energy for transmitting some (parts or all) remaining buffer data.
The reasons are as follows.
First, using overflowed energy for transmitting some (parts or all) remaining buffer data will not increase one-step cost since no extra grid power is
utilized.
Second, using overflowed energy for transmitting some (parts or all) remaining buffer data will decrease the initial buffer data for future while the
initial battery energy for future does not change (remains $E_{max}$). Using Property \ref{Increasing in q} in Appendix \ref{Properties of Valpha}, we
derive that the average cost will be decreased.
\end{IEEEproof}
\emph{Remark: Lemma \ref{Sufficient condition for the non-optimality} means if a policy results in battery overflow but non-emptiness of the
buffer, there are (is) polices (policy) that can achieve better average cost performance definitely.
}
\par
\emph{Remark: Lemma \ref{Sufficient condition for the non-optimality} gives a sufficient condition for the non-optimality. Meanwhile,
Lemma \ref{Sufficient condition for the non-optimality} can be also viewed as the necessary condition for the optimality. That is to say, any
average optimal policy should not incur battery overflow and non-emptiness of the buffer simultaneously.
}

Next, based on Lemma \ref{discout to UP} and Proposition \ref{Necessary condition for the discout optimality} in Appendix \ref{discount optimal}, we
have the following lemma.
\begin{lemma}\label{Necessary condition for the beta optimality}
 Given state $x=(q,h,a,e_b,e)$, the average cost optimal policy $(r^*(x),w^*(x))$
 should satisfy the following inequality array
\begin{eqnarray}\label{u beta}
\tilde{Z}_1(q,q-r^*,h,a,e_b-w^*,e) \le \beta\rho \frac{\sigma^2}{h}e^{\theta q}(e^{\theta}-1)
 \le \tilde{Z}_1(q,q-r^*+1,h,a,e_b-w^*,e),
\end{eqnarray}
\begin{eqnarray}\label{eta beta}
\tilde{Z}_2(q-r^*,h,a,e_b-w^*,e)\le \frac{-\beta}{\tau}
\le \tilde{Z}_2(q-r^*,h,a,e_b-w^*+1,e),
\end{eqnarray}
\begin{eqnarray}\label{u eta beta}
\tilde{Z}_3(q,q-r^*,h,a,e_b-w^*,e) \le \beta\rho \frac{\sigma^2}{h}e^{\theta q}(e^{\theta}-1)
 \le \tilde{Z}_3(q,q-r^*+1,h,a,e_b-w^*+1,e),
\end{eqnarray}
where $\tilde{Z}_1(q,u,h,a,\eta,e)=\lim\limits_{\alpha \to 1}Z_1(q,u,h,a,\eta,e)$, $\tilde{Z}_2(u,h,a,\eta,e)=\lim\limits_{\alpha \to
1}Z_2(u,h,a,\eta,e)$, and $\tilde{Z}_3(q,u,h,a,\eta,e)=\lim\limits_{\alpha \to 1}Z_3(q,u,h,a,\eta,e)$. $Z_i(\cdot)$ ($i=1,2,3$) is defined in
Proposition \ref{Necessary condition for the discout optimality}.
\end{lemma}

\emph{Remark:
Lemma \ref{Necessary condition for the beta optimality} reveals a necessary condition for the average cost optimality, i.e., the optimal transmit rate $r^*$ and the optimal battery energy allocation $w^*$ should satisfy the condition.
}
\par
\emph{Remark:
When $(r^*,w^*)$ is on the boundary of the feasible set, corresponding conditions can also be obtained similarly.
}
\par
Combining Lemma \ref{discout to UP} and Proposition \ref{discount optimal for two special cases} in Appendix \ref{discount optimal}, we derive the
following lemma.
\begin{lemma}\label{average optimal for two special cases}
For $x=(q,h,a,e_b,e)$ satisfying
\begin{eqnarray}\label{5}
\tilde{Z}_1(q,0,h,a,\tau\max\{0,\frac{e_b}{\tau}-P(x,q)\},e)
> \beta\rho \frac{\sigma^2}{h}e^{\theta q}(e^{\theta}-1)
\end{eqnarray}
and
\begin{eqnarray}\label{6}
\tilde{Z}_2(0,h,a,\tau\max\{0,\frac{e_b}{\tau}-P(x,q)\},e)> \frac{-\beta}{\tau},
\end{eqnarray}
$(q,e_b-\tau\max\{0,\frac{e_b}{\tau}-P(x,q)\})$ is the average cost optimal policy.
In addition, for $(q,h,a,e_b,e)$ satisfying
\begin{eqnarray}\label{7}
\tilde{Z}_1(q,q,h,a,e_b,e)< \beta\rho \frac{\sigma^2}{h}e^{\theta q}(e^{\theta}-1)
\end{eqnarray}
and
\begin{eqnarray}\label{8}
\tilde{Z}_2(q,h,a,e_b,e)< \frac{-\beta}{\tau},
\end{eqnarray}
$(0,0)$ is the average cost optimal policy.
\end{lemma}

\emph{Remark:
$(q,e_b-\tau\max\{0,\frac{e_b}{\tau}-P(x,q)\})$ means
transmit all the data in the buffer and the allocated battery energy
is $e_b-\tau\max\{0,\frac{e_b}{\tau}-P(x,q)\}$). That is to say, transmit all data in the buffer and allocate as much energy as possible from the
battery. Specifically, if the required power for transmitting all buffer data is less than the power stored in the battery, allocate all the required
power from the battery. Otherwise, allocate all the battery's energy (the rest of the required power will be allocated from the power grid).
$(0,0)$ means  transmit no buffer data and  allocate no battery energy.
}

\emph{Remark: (\ref{5}) and (\ref{6}) give the set of states, for which transmit all the buffer data the together with allocate as much energy as
possible from the battery is the two-dimensional average cost optimal policy. (\ref{7}) and (\ref{8}) give the set of states, for which transmit no
buffer data together with allocate no battery energy is the two-dimensional average cost optimal policy.
}

In the following, we investigate the monotonicity.

\begin{lemma} \label{non-increasing in beta}
Denote the optimal stationary deterministic policy for UP$_\beta$ as $g_\beta$, we have
\begin{itemize}[]
\item $J^{g_\beta}(\beta)$ is non-decreasing in $\beta$.
\item $B^{g_\beta}$ is non-decreasing in $\beta$, and $K^{g_\beta}$ is monotone non-increasing in $\beta$
\end{itemize}
\end{lemma}
 \begin{IEEEproof}
See Appendix \ref{Proof of lemma non-increasing in beta}.
    \end{IEEEproof}

  \begin{lemma} \label{monot of optimal policy}
  The average cost optimal transmit rate policy $r(q,h,a,e_b,e)$ is non-decreasing in $q$ and $e_b$, respectively; The average cost optimal battery energy allocation policy
  $w(q,h,a,e_b,e)$ is non-decreasing in $q$ and $e_b$, respectively.
 \end{lemma}
  \begin{IEEEproof}
  The lemma can be proved by the second half of Lemma \ref{discout to UP} and Proposition \ref{monot of discount optimal policy} in Appendix \ref{discount optimal}.
  \end{IEEEproof}


\section{Relations between rate allocation and the battery power allocation}\label{dimesion reduction}
\begin{figure*}[!t]
\centering
\includegraphics[width=4.0in,angle=90]{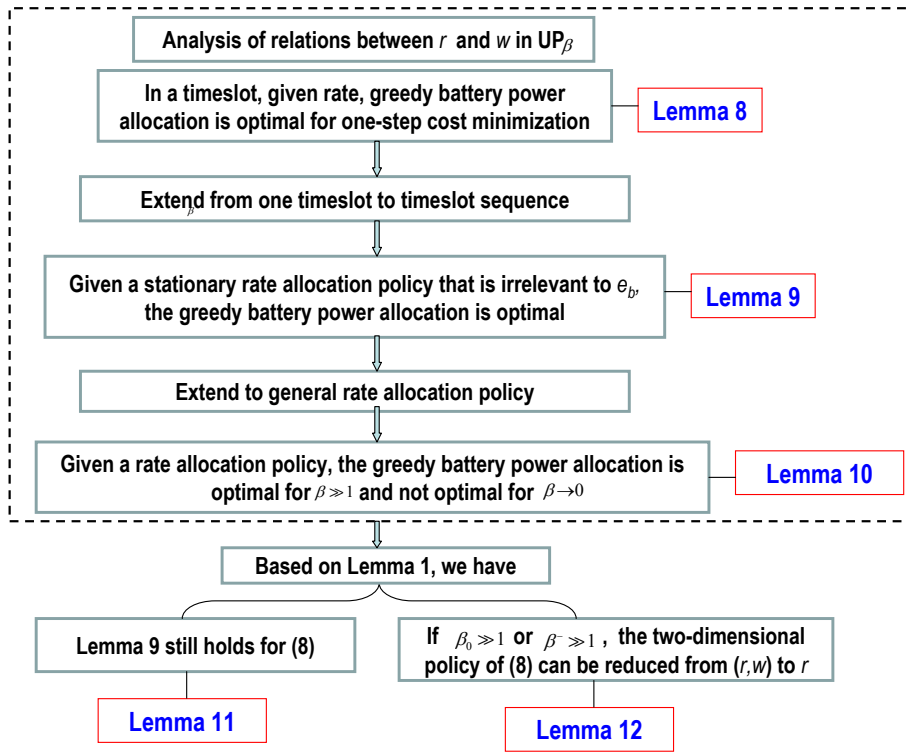}
\caption{Analysis structure of Section \ref{dimesion reduction}, the derived results are also shown correspondingly}
\label{Dimensionreducation}
\end{figure*}

The rate allocation $r$ and the power allocation from the battery $w$ are coupled together, they affect each other.
In this section, we investigate the relations between the rate allocation $r$ and the power allocation from the battery $w$.
We first focus on the relation between $r$ and $w$ in UP$_\beta$. Next, we derive that under a condition, the policy of the constrained
two-dimensional MDP problem (\ref{original MDP}) can be reduced to the rate policy only.
To make the presentation clear, the analysis structure of this section is drawn in Fig. \ref{Dimensionreducation}.
\subsection{The relation between $r$ and $w$ in UP$_\beta$}
If we assume that rate $r[n]$ has been chosen at the $n$-th time-slot, then the required total power has been fixed. In this case, to minimize the
immediate one-step cost $q[n]+\beta\big[\rho \frac{\sigma^2}{h[n]}(e^{\theta r[n]}-1)+\Delta(r[n])-w[n]\big]^{+}$, we will allocate as much power as
possible from the battery to meet the required total power, i.e., the greedy policy for the battery power allocation. This is because the power from
the battery is \lq\lq free\rq\rq.\footnote{Please refer to (\ref{UP_beta}). the price of the grid power is $\beta$. } Formally, we have the following
claim.
\begin{lemma}\label{one-stepcost}
In a timeslot, if the rate allocation $r$ is chosen, the greedy battery power allocation is optimal for the immediate one-step cost minimization.
\end{lemma}
\begin{IEEEproof}
See Appendix \ref{Poof of one-stepcost}.
\end{IEEEproof}
\emph{Remark: Lemma \ref{one-stepcost} reveals the optimality of greedy battery power allocation for given rate in a time-slot.}

In the following, we consider the extension from one time-slot to the timeslot sequence.
First, we have the following lemma.
\begin{lemma}\label{optimalforgreedy}
For a given rate allocation policy $r(q,h,a,e_b,e)$ that is irrelevant to $e_b$, i.e., $r(q,h,a)$,\footnote{According to (\ref{energy evolution in the
battery}), if a policy is irrelevant to $e_b$, then it is irrelevant to $e$.} the greedy battery power allocation is optimal (for UP$_\beta$).
\end{lemma}
\begin{IEEEproof}
See Appendix \ref{Poof of optimalforgreedy}.
\end{IEEEproof}
\emph{Remark: The irrelevance to $e_b$ of rate allocation policy is sufficient condition for the optimality of greedy battery power allocation. Lemma
\ref{optimalforgreedy} guarantees the optimality of greedy battery power under any given rate allocation policy irrelevant to $e_b$. }
\par
Next, a natural question is \emph{whether the greedy allocation strategy of battery power is optimal given general rate allocation
policy $r(x=(q,h,a,e_b,e))$?} The following lemma gives the answer.
\begin{lemma}\label{lemma greedy optimality discussion}
Given a rate allocation policy $r(x)$,\footnote{According to Lemma \ref{monot of optimal policy}, it is reasonable to assume that $r(x)$ is
non-decreasing in $q$ and $e_b$, respectively. }
\begin{itemize}[]
\item When $\beta$ is large enough, e.g., $\beta \gg 1$, the greedy policy is the optimal battery power allocation policy in UP$_\beta$.
\item If $\beta$ is sufficiently small, e.g., $\beta \to 0$, the greedy battery power allocation policy is NOT optimal for UP$_\beta$.
\end{itemize}
\end{lemma}
\begin{IEEEproof}
See Appendix \ref{Proof of lemma greedy optimality discussion}.
\end{IEEEproof}
\par
\emph{Remark: Lemma \ref{lemma greedy optimality discussion} reveals that the greedy policy is NOT the optimal battery power allocation policy in
UP$_\beta$ with arbitrary $\beta$. The optimality of greedy battery power allocation depends on the value of $\beta$. It can be explained as follows:
Since $\beta$ is the \lq\lq price\rq\rq~of grid power in UP$_\beta$, when the grid power is very cheap, the profit of reserving some battery power for future timeslot\footnote{Based on Property \ref{Non-increasing in eb} in Appendix \ref{Properties of Valpha}, there exists profit for reserving some
battery power for future timeslot. The price of using grid energy is constant over time, the cost of using grid power is constant. But reserving battery energy can incur more data transmission in future (Observe that the rate policy has been given already, more battery power leads to more data transmission). That is to say, delaying the use of battery energy has profits in minimizing data delay. All in all, there are profits for the first part of $J_x^{\pi}(\beta)$.} is more than the cost of buying the same amount of grid power in current timeslot. Thus, reserving some battery energy but using the grid power instead is optimal. When the price is high, the cost of buying the grid power is more than the profit of reserving some battery energy, then allocate as much energy as possible from the battery to fulfill the required power (i.e., greedy battery allocation policy) is optimal.}
\par
\emph{Remark:
 As the remaining battery energy will affect action and cost in future timeslot for given rate policy (e.g., battery power allocation $w[n]$ at the $n$-th time-slot will
 affect the rate allocation $r[n+1]$ at the $(n+1)$ times-lot),
 the optimality of greedy battery power allocation can not extend from one timeslot (Lemma \ref{one-stepcost}) to time-slot sequence.
 }

\subsection{Dimension reduction for the two-dimensional policy of the constrained MDP under a sufficient condition}
According to Lemma \ref{OP of the general}, the two-dimensional optimal policy of constrained MDP (\ref{original MDP})
can be derived by the optimal policy of the UP$_{\beta}$ with one or two values of $\beta$. Then we have
\begin{lemma}
For a given rate allocation policy $r(q,h,a,e_b,e)$ that is irrelevant to $e_b$, i.e., $r(q,h,a)$, the greedy battery power allocation is optimal
(for the constrained MDP (\ref{original MDP})).
\end{lemma}
Furthermore,
the following lemma reveals that the two-dimensional policy of the constrained MDP can be reduced to the rate policy when $\beta_0$ or $\beta^-$
satisfies a condition.
\par
\begin{lemma}\label{dimension reduction under a condition}
If $\beta_0\gg 1$ or $\beta^-\gg 1$, the greedy policy is the optimal battery power allocation policy of the two-dimensional constrained MDP
(\ref{original MDP}). Furthermore,
view $(X[n],R[n])$ as an MDP with state $X[n]$ and action $R[n]$.\footnote{The state includes the buffer queue length, channel gain, data arrival,
energy in the battery, and harvested energy arrival. The action includes the allocated rate only.} The feasible action $r$ in state $x=(q,h,a,e_b,e)$
belongs to $\{0,1,\cdots,q\}$.
Define $\pi_r=(\pi_r[0],\pi_r[1],\cdots)$ to be a policy that $\pi_r[n]$ generates an action $r[n]$ at $n\tau$, the optimal policy of the following
MDP problem is the optimal rate policy of (\ref{original MDP}).
\begin{eqnarray} \label{original optimization problem converted to policy only r}
\min_{\pi_r}\mathop {\lim \sup }\limits_{n\to \infty} \frac{1}{n}\mathbb{E}_{x}^{\pi_r}\left[\sum\limits_{k =0}^{n-1} {Q[k]}\right]
\end{eqnarray}
\begin{equation}
\mbox{s.t.}\quad \mathop {\lim \sup }\limits_{n\to \infty} \frac{1}{n}\mathbb{E}_{x}^{\pi_r}\left[\sum\limits_{k =0}^{n-1} P_{grid}[k]\right] \le
\bar{\mathcal{P}},
\end{equation}
where
$
P_{grid}[k]=
P(X[k],R[k])-\min\left\{P\Big(X[k],R[k]\Big),\frac{E_{b}[k]}{\tau}\right\},
$
and the evolution of energy in the battery becomes
$
E_{b}[k+1]=
\left(E_b[k]-\tau\min\left\{P\Big(X[k],R[k]\Big),\frac{E_{b}[k]}{\tau}\right\}+E[k]\right)^{-}.
\nonumber
$
\end{lemma}
\begin{IEEEproof}
See Appendix \ref{Proof of dimension reduction under a condition}.
\end{IEEEproof}

\emph{Remark:
When the condition $\beta_0 \gg 1$ or $\beta^-\gg 1$ holds, the two-dimensional policy can be obtained as follows. We can first derive the optimal battery
allocation policy of the two-dimensional policy to be greedy policy, and then the optimal rate policy can be solved through an MDP whose policy
includes the rate allocation only (i.e., (\ref{original optimization problem converted to policy only r})). The dimension of the policy has reduced
from $(r,w)$ to $r$.
}

\emph{Remark: If $\beta_0 \gg 1$ or $\beta^-\gg 1$, the dimension reduction can be implemented. In contrast, if $\beta_0 \to 0$ or $\beta^+ \to 0$,
the dimension reduction in Lemma \ref{dimension reduction under a condition} can not be accomplished (See the second half of Lemma \ref{lemma greedy
optimality discussion}). For other cases, we do not know whether the dimension reduction can be implemented.
$\beta_0 \gg 1$ or $\beta^-\gg 1$ is only a sufficient condition for dimension reduction in Lemma \ref{dimension reduction under a condition}.
}
\par
 Since there is a condition $\beta_0 \gg 1$ or $\beta^-\gg 1$ in Lemma \ref{dimension reduction under a condition} and the dimension reduction does
 not hold for $\beta_0 \to 0$ or $\beta^+\to 0$, formulating the original optimization problem (\ref{original optimization problem}) directly as (\ref{original
 optimization problem converted to policy only r}) is NOT convincing.\footnote{If we can prove that the condition $\beta_0 \gg 1$ or $\beta^-\gg 1$ holds,
 (\ref{original optimization problem}) can be reformulated as (\ref{original optimization problem converted to policy only r}).}
\section{Policy of the constrained MDP}\label{Proposedpolicies}
\begin{figure}[!t]
\centering
\includegraphics[width=4.0in,angle=90]{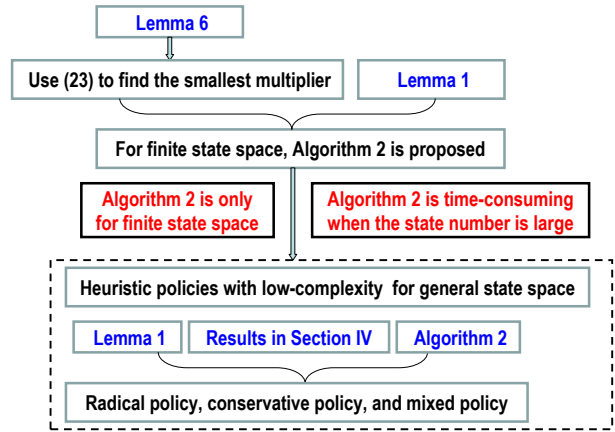}
\caption{Structure of Section \ref{Proposedpolicies}}
\label{constrainedpolicy}
\end{figure}
Based on the previous theoretical results, an algorithm to find the constrained optimal policy is proposed for the finite state space, and heuristic polices are given for the general state space. The structure of this section is illustrated in Fig. \ref{constrainedpolicy}.
\subsection{Algorithm to find the optimal policy for finite state space}
In this subsection, we give the algorithm to find the constrained optimal policy when the state is finite.\par According to Lemma \ref{non-increasing
in beta}, smaller $\beta$ results in better delay performance $B$.
Meanwhile, the decrease of $\beta$ will increase the grid power consumption $K$. Too small $\beta$ will violate the grid power constraint.
Then
we should find the smallest $\beta$ that satisfying the average grid power first. Denote
$\beta^*=\mathrm{inf}\{\beta:K^{g_\beta} \le \bar{\mathcal{P}}\}$, where $g_\beta$ is the optimal policy of UP$_\beta$. We can use the following
method to find $\beta^*$. Let
\begin{eqnarray}\label{finding inf beta}
\beta_{n+1}=\beta_{n}+\frac{1}{n}\big(K^{g_{\beta_n}}-\bar{\mathcal{P}}\big)
\end{eqnarray}
with $\beta_1$ is a sufficiently large number. $K^{g_{\beta_n}}$ is computed by using the relative value iteration algorithm for each $\beta_n$.
Then $\{\beta_n\}$ converges to $\beta^*$ \cite{IMA07:D. V. Djonin and V. Krishnamurthy}. Based on Lemma \ref{OP of the general}, if the average grid
power $K^{g_{\beta^*}}$ equals to the grid power constraint, the obtained optimal policy is also optimal for the constrained MDP. Otherwise, we should find $\beta^+$ and $\beta^-$. The detailed algorithm for
finite state is listed in Table \ref{Algorithm of finding the constrained optimal policy}.

 \begin{table}[]
 \caption{\label{Algorithm of finding the constrained optimal policy}}
 \centering
 \begin{tabular}{lcl}
  \toprule
  \textbf{Algorithm 2: Algorithm of finding the constrained optimal policy for finite state} \\
  \midrule
 Step 1:  \\
 Using iteration algorithm (\ref{finding inf beta}) to find $\beta^*$, and the corresponding average grid \\
 power $K^{g_{\beta^*}}$, in which the relative value iteration algorithm (Algorithm 1) is applied. \\
 Step 2:   \\
 If $K^{g_{\beta^*}}=\bar{\mathcal{P}}$, then $g_{\beta^*}$ is the optimal policy of the constrained MDP.
 Otherwise, go to next Step.\\
 Step 3:   \\
 Perturb $\beta^*$ by $\nu$: $\beta^+=\beta^*+\nu$ and $\beta^-=\beta^*-\nu$. Find the optimal policies $g_{\beta^+}$ and $g_{\beta^-}$ for \\
 UP$_{\beta^+}$ and UP$_{\beta^+}$ as well as the corresponding grid power $K^{g_{\beta^+}}$ and $K^{g_{\beta^-}}$, respectively, by using \\
Algorithm 1. The optimal policy is taking $g_{\beta^+}$ with probability $\xi$ and $g_{\beta^-}$ with probability   \\
$1-\xi$ at each decision stage. $\xi$ is determined by $\xi K^{g_{\beta^+}}+(1-\xi)K^{g_{\beta^-}}=\bar{\mathcal{P}}$.\\
    \bottomrule
 \end{tabular}
\end{table}
\subsection{Proposed heuristic policies}
Algorithm 2 is only for the finite state space. Meanwhile, it is time-consuming when the number of states is large.
In this subsection, we propose low-complex heuristic policies for general state space.
The paper has derived the structural properties of the optimal policy. Particularly,
we have proved that the optimal policy exists, and it is a stationary deterministic policy or a mixed policy of two stationary deterministic policies.
Moreover, we have proved that the greedy battery power allocation MAY BE optimal (in Section \ref{dimesion reduction}).
Based on these properties and in light of Algorithm 2,
we propose heuristic policies as follows (a summary is given in Table \ref{stationary deterministic policy}).
\par
The first is named radical policy. Under radical policy, the action is $(r=q,w=\min\{e_b,P(x,r)\})$ for state $x=(q,h,a,e_b,e)$.
That is to say, all the buffer data are served at each time-slot, and use the greedy strategy for the
battery energy allocation, i.e., if the required power is not greater than the battery power, then all the power will be supplied from the battery and
no grid power will be used. Otherwise, allocate all the battery power, and the rest will be supplied from the power grid.
\par
\emph{Remark: When there is no average grid power constraint, the radical policy is the optimal policy to minimize the mean buffer delay. Furthermore,
given an average grid power constraint, when the mean date arrival, mean energy arrival, and mean channel gain satisfy a condition, the grid power
constraint can be obeyed under radical policy, the radical policy is the optimal policy even when considering the average grid power constraint.
}
\par
In the radical policy, the average grid power constraint is not
considered. Then we propose another policy (i.e., the conservative policy) that guarantees the average grid power constraint
through satisfying the constraints in each time-slot. Define $P^{-1}(\cdot)$ as the inverse function of $P(x,r)$ with respect to $r$.
We call the policy $(r(x),w(x))= \left(\min\left\{q,P^{-1}\left(\bar{P}+\frac{e_b}{\tau}\right)\right\},\min\{\frac{e_b}{\tau},P(x,r)\}\right)$ the
conservative policy. That is to say, we first guarantee that the
grid power utilized in each time is less than the average grid
constraint, then transmit as many packets as possible and utilize
the greedy policy for the battery energy allocation.
\par
The third policy is a random policy referred to as mixed policy. In the mixed policy, the radical policy and conservative policy are utilized randomly with probability $\xi$ and $1-\xi$, respectively. Denote the average grid power consumptions of the radical policy and conservative as $G_r$ and $G_c$, respectively. $\xi$ is determined by $\xi*G_r+(1-\xi)*G_c=\bar{\mathcal{P}}$.

 \begin{table*}[]
 \caption{\label{stationary deterministic policy}}
 \centering
 \begin{tabular}{llll}
  \toprule
Policy name & Strategy $(r(x),w(x))$ for $x=(q,h,a,e_b,e)$ \\
  \midrule
  Radical policy &$\left(q,\min\{e_b,P(x,r)\}\right)$  \\
 Conservative policy & $\left(\min\left\{q,P^{-1}\left(\bar{P}+\frac{e_b}{\tau}\right)\right\},\min\{\frac{e_b}{\tau},P(x,r)\}\right)$ \\
Mixed policy & Apply the radical policy and conservative policy with  \\
& probability $\xi$ and $1-\xi$, respectively  \\ \bottomrule
 \end{tabular}
\end{table*}

\section{Numerical results}
In this section, simulation results are presented under the radical policy, conservative policy and mixed policy. We consider the i.i.d. Rayleigh fading channel (i.e., the power gain $H$ is exponentially
distributed). In addition,
unless otherwise specified, we set $\tau=1$, $b=1$, $N=5$, and $\rho=1$. Both the initial battery energy and initial buffer length are zero.
\par
\begin{figure}[]
\centering
\includegraphics[width=2.5in]{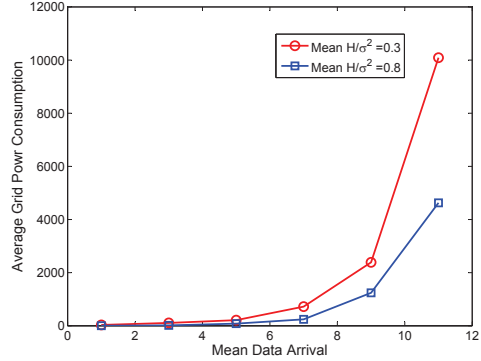}
\vspace{-1em}
\caption{Average grid power consumptions v.s. $\bar{A}$. $C=1$ and $E_{max}=2500$.
$A$ takes $0$ and $2*\bar{A}$ with equal probability $0.5$. $E$ takes values $\{200, 800, 1000, 2000\}$ with probabilities $\{0.1, 0.6, 0.2, 0.1\}$,
respectively. The mean grid power is average over $10^6$ time-slots.}
\label{grid power vs data arrival}
\vspace{-1em}
\end{figure}

Fig. \ref{grid power vs data arrival} plots the average grid power consumption with respect to the average data arrival ($\bar{A}$) under radical
policy.
We can observe that when $\bar{A}$ is small, the grid power consumption is nearly zero. However, when $\bar{A}$ is large
the grid power consumption grows rapidly with the increase of $\bar{A}$ roughly according to exponential relation.
This can be explained as follows: when $\bar{A}$ is small, the required power is small and the battery can supply the power. Then no grid power will
be consumed. Once $\bar{A}$ is large, the required power is much larger than the battery power, and the grid power becomes the main power source.
Since the required power roughly varies with the transmission rate according to the exponential function, the grid power consumption varies
exponentially with $\bar{A}$. Meanwhile, we can see that the better channel conditions lead to less grid power consumption.
\par
Furthermore, from Fig. \ref{grid power vs data arrival}, it can be derived that if $\bar{A}$ is less than a certain value, the grid power will be less
than a certain value.
 Since the radical policy is optimal for the buffer delay minimization without the average grid power constraint, if $\bar{A}$ is less than some value
 to make the average grid power be no more than the constraint, i.e., the average grid power constant is satisfied, then the radical policy is also
 optimal when considering the grid power constraint. For example, when $\bar{\mathcal{P}}=2000$, according to Fig. \ref{grid power vs data arrival}, the
 strategy is optimal for $\bar{A}=1,2,\cdots,8$.
The reason is that when the average power grid plus the harvested power is large enough to serve all the data, then serving all is optimal.
\par
\begin{figure}[]
\centering
\subfigure[v.s. $\bar{\mathcal{P}}$. $\overline{H/\sigma^2}=0.3$]{\includegraphics[width=2.5in]{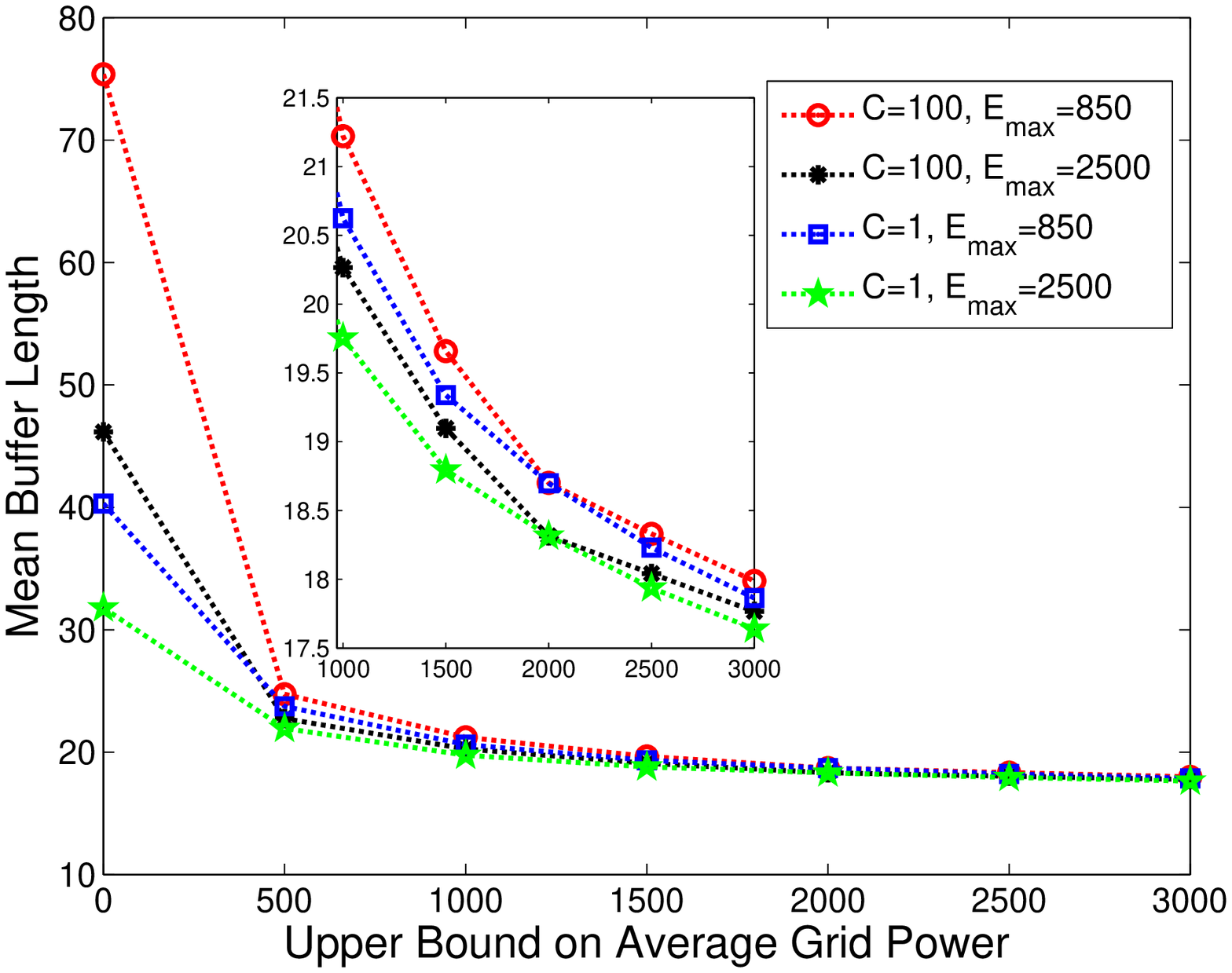}\label{barP Vs mean
buffer under conservative}}
\subfigure[v.s. $\overline{H/\sigma^2}$. $\bar{\mathcal{P}}=1000$]{\includegraphics[width=2.5in]{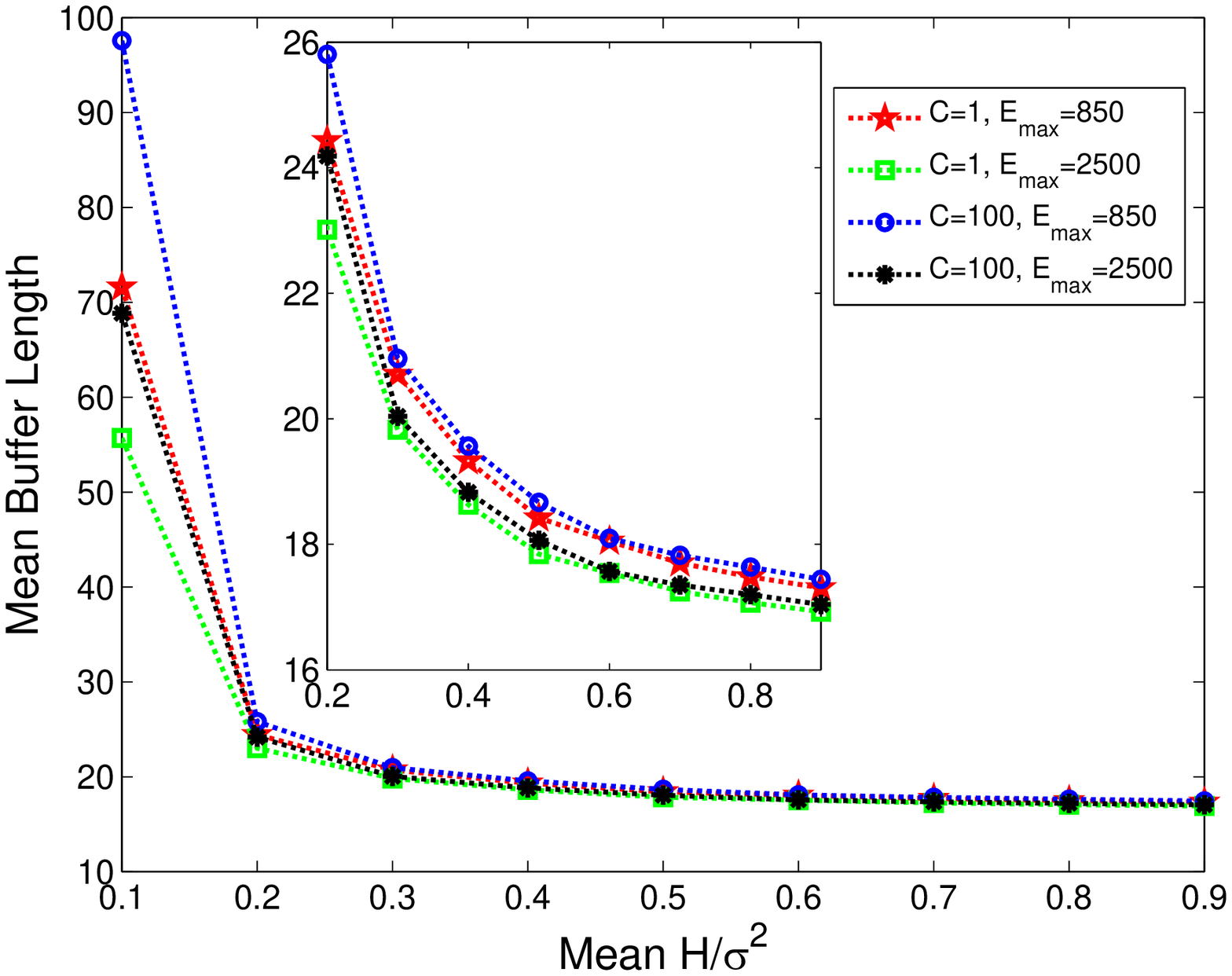}\label{MeanH Vs mean buffer under conservative}}
\caption{The mean buffer length performance for conservative policy.}
\label{Conservative policy}
\vspace{-1em}
\end{figure}

Fig. \ref{Conservative policy} illustrates the average buffer length performance
for conservative policy. $A$ takes values from $\{0,10,20,30\}$ with probabilities $\{0.1, 0.3, 0.5, 0.1\}$,
respectively. $E$ takes values $\{200, 800, 1000, 2000\}$ with probabilities $\{0.1, 0.6, 0.2, 0.1\}$, respectively.
In Fig. \ref{barP Vs mean buffer under conservative}, the buffer length is averaged over $10^5$
time-slots.
From the figure, we can see that the mean buffer length decreases fast when $\bar{\mathcal{P}}$ is small (e.g., $\bar{\mathcal{P}}\le 1000$ ), and the decrease becomes
slow when $\bar{\mathcal{P}}$ is large (e.g., $\bar{\mathcal{P}}>2000$). This can be explained as follows: when the upper bound of the average grid power (i.e.,
$\bar{\mathcal{P}}$) increases, there are more available grid power in a time-slot in average, sense and we can transmit more (at least no less) buffer data,
then the average buffer length becomes shorter. When $\bar{\mathcal{P}}$ is small,
$r=\min\left\{q,P^{-1}\left(\bar{\mathcal{P}}+\frac{e_b}{\tau}\right)\right\}=P^{-1}\left(\bar{\mathcal{P}}+\frac{e_b}{\tau}\right)$ with a high chance, hence $r$
increases apparently with the increase of $\bar{\mathcal{P}}$, and the average buffer length decreases quickly. Once $\bar{\mathcal{P}}$ is large enough,
$r=\min\left\{q,P^{-1}\left(\bar{\mathcal{P}}+\frac{e_b}{\tau}\right)\right\}=q$ with a high probability, and $r$ becomes static with respect to $\bar{\mathcal{P}}$.
Then, the average buffer length decreases slowly. Furthermore, we can observe that more extra circuit power consumption (i.e., $C$) and smaller
battery capacity can respectively result in worse mean buffer length performance (i.e., longer length). Meanwhile,
by comparing $(C=1, E_{max}=850)$ with $(C=100, E_{max}=2500)$, we can find that the mean buffer length performance for $(C=1, E_{max}=850)$ is better
when $\bar{\mathcal{P}}$ is small. But when $\bar{\mathcal{P}}$ is large, $(C=100,E_{max}=2500)$ has slightly better performance.
\par
In Fig. \ref{MeanH Vs mean buffer under conservative}, for each curve, we can observe
that the buffer length performance decreases with the increase of $\overline{H/\sigma^2}$, fast when $\overline{H/\sigma^2}$ is small (e.g.
$0.1,0.2,0.3$), moderately when $\overline{H/\sigma^2}$ is large (e.g., $0.4,\cdots,0.7$), and slowly when $\overline{H/\sigma^2}$ is very large
(e.g., $0.8,0.9$).
The reason is as follows.
When $\overline{H/\sigma^2}$ is not very large, $q>P^{-1}(\frac{e_b}{\tau})$ with a high probability, i.e., $r(x)=P^{-1}(\frac{e_b}{\tau})$. The
remaining buffer length $u(x)=q-r(x)=q-P^{-1}(\frac{e_b}{\tau})$ will decrease with the increase of $h/\sigma^2$ approximately according to minus
logarithmic relation.\footnote{$P^{-1}(\cdot)$ is increasing with $h/\sigma^2$ according to logarithmic relation.} Thus, the mean buffer length
decreases harshly at first and moderate then.
Once $\overline{H/\sigma^2}$ is larger than a certain value, $q<P^{-1}(\frac{e_b}{\tau})$ with a high probability. Then, $r(x)=q$ and the remaining
buffer length becomes zero with a high probability. In this case, the increase of $\overline{H/\sigma^2}$
will not have great effects on the mean buffer length (the mean length is nearly the average data arrival, $16$).
\par
\begin{figure}[]
\centering
\includegraphics[width=2.5in]{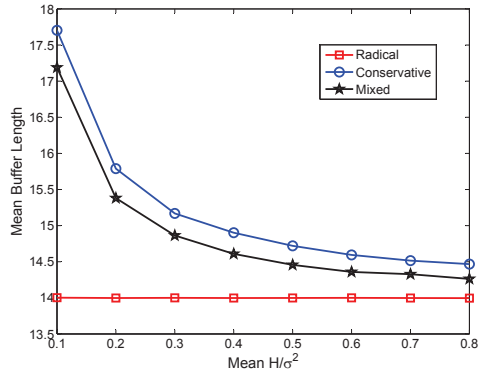}
\vspace{-1em}
\caption{Buffer length performance of the radical policy, conservative policy, and mixed policy}
\label{BufferlengthCompare}
\vspace{-1em}
\end{figure}
\begin{figure}[]
\centering
\subfigure[Radical policy]{\includegraphics[width=2in]{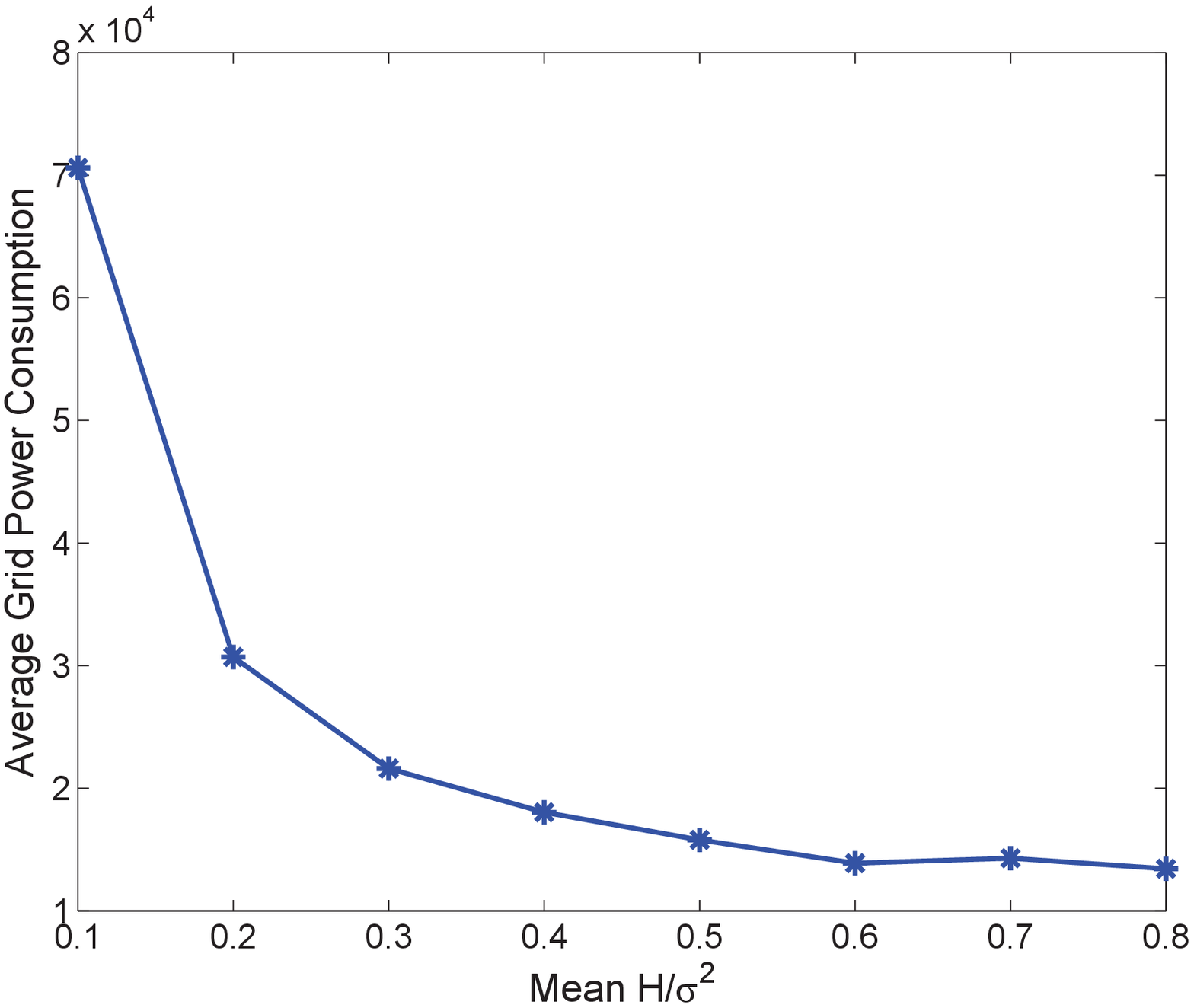}\label{}}
\subfigure[Conservative policy]{\includegraphics[width=2in]{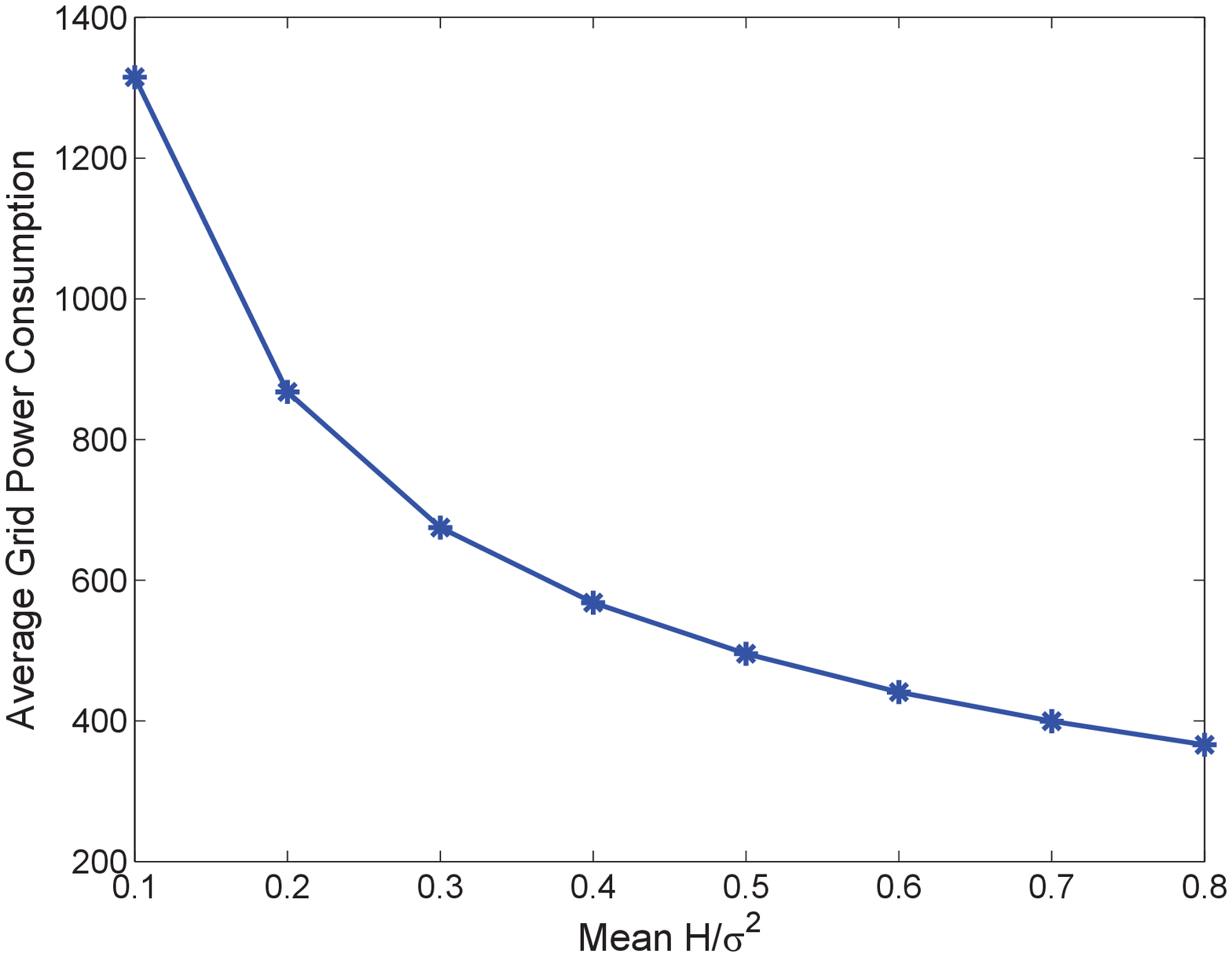}\label{}}
\caption{The average grid power consumptions of the radical policy and conservative policy.}
\label{Grid power for RadicalConservative policy}
\vspace{-1em}
\end{figure}

Fig. \ref{BufferlengthCompare} compares the buffer length performance of the heuristic policies with respect to $\bar{H}/\sigma^2$.
In the simulations, $A$ takes values from $\{0,10,20,30\}$ with probabilities $\{0.1, 0.5, 0.3, 0.1\}$,
respectively. $E$ takes values $\{200, 800, 1000, 2000\}$ with probabilities $\{0.1, 0.6, 0.2, 0.1\}$, respectively.. $E_{max}=2500$ and
$\bar{\mathcal{P}}=3000$. Based on the average grid power consumptions of radical policy and conservative policy (as plotted in Fig. \ref{Grid power for RadicalConservative policy}), we compute the probability of using radical policy in the mixed policy, $\xi=[0.9468~    0.8615~    0.7933~    0.7463~    0.7053~    0.6608~    0.6689    ~0.6452]$. We can see that in terms of the buffer length performance, the radical policy is better than the mixed policy, which is better than the conservative policy.
For the conservative policy and mixed policy, the buffer length decreases with the increase of $\bar{H}/\sigma^2$ first harshly and then moderately.
The explanations for the conservative policy are similar to Fig. \ref{MeanH Vs mean buffer under conservative}. As the usage probability of the conservative policy in mixed policy is high, the buffer length of the mixed policy is similar as the conservative policy. Meanwhile, as there is chance of using the radical policy in the mixed policy, the buffer length performance of the mixed policy is better than the conservative policy.
The mean buffer length of the radical policy is approximately the mean data arrival and remains static. As the radical policy is the optimal policy without the grid power constraint, the buffer length of the radical policy is the lower bound of the optimal policy.

\section{Conclusion}
In this paper, we have studied the power allocation of the physical layer together with the optimal mean buffer delay of the upper layer in green
networks with energy harvesting nodes. The physical power allocation contains two aspects: power allocation from the power grid and power allocation
from the battery. The rate allocation can represent the total power allocation and the grid power allocation is the total power subtract the battery
power, then the physical power allocation is equivalent to rate allocation and battery power allocation.
For the purpose of modeling and analyzing the conflicting relation between power and delay as well as the coupling between rate allocation and battery power allocation, we reformulate a constrained MDP with a two-dimensional policy. The analysis of the constrained MDP is transformed to that of the corresponding unconstrained MDP. Structural properties of the optimal policy are derived. In addition, the relations between elements of the two-dimensional policy are also investigated. According to the theoretical study, an algorithm to find the constrained optimal policy is proposed for finite state space. Furthermore, heuristic policies (i.e., the radical policy, the conservative policy and the mixed policy) are presented for general state.
In the end, simulations are carried out under these policies. We have observed the interactions among the channel, the data
arrival, the harvested energy arrival,  the power grid, and the data buffer length.

\linespread{1.2}

\appendix
\subsection{Proof of Lemma \ref{OP of the general}}\label{Proof of Lemma OP of the general}
If for some $\beta$ (denoted as $\beta_0$), the optimal stationary policy $\pi^*$ of UP$_{\beta_0}$ satisfies: 1) $\pi^*$ yields $B^{\pi^*}$ and
 $K^{\pi^*}$ as limits for all $x \in \mathcal{X}$;
2) $K^{\pi^*}=\bar{\mathcal{P}}$. Then $\pi^*$ is optimal for the constrained MDP (\ref{original MDP}) according to \cite{CDC86:D. J. Ma A. M. Makowski and A.
Shwartz}\cite{JMAA85:F. J. Beutlerand and K. W. Ross}. Otherwise, there are $\beta^+$ and $\beta^-$. The optimal policy $\pi^-$ that obtained for
UP$_{\beta^-}$ has a grid power consumption slightly larger than $\bar{\mathcal{P}}$. $\beta^+ > \beta^-$ will instead lead to a less aggressive
policy $\pi^+$ with a grid power consumption slightly smaller than $\bar{\mathcal{P}}$. The optimal policy for the constrained MDP (\ref{original
MDP}) is as follows: at each decision epoch, choose $\pi^-$ with a certain probability $q$ and $\pi^+$ with probability $1-q$, where $q$ depends on
$\bar{\mathcal{P}}$ and the grid power consumptions of the two policies \cite{JMAA85:F. J. Beutlerand and K. W. Ross}\cite{PEIS93:L. I. Sennott}.\footnote{The state space is countable.}
\subsection{Proof of Lemma \ref{discout to UP}}\label{Proof of Lemma discout to UP}
We prove the lemma by applying Theorem 3.8 in \cite{MOR93:M. Schal}.
First, we can prove that the conditions of Proposition 2.1 in \cite{MOR93:M. Schal} holds. Next,
the discounted cost optimality equation \cite{Book96:O. H. Lerma and J. B. Lassere} for $V_{\alpha}(x)$ is
\begin{eqnarray}\label{DCOE}
  V_{\alpha}(q,h,a,e_b,e)=\min_{r\in \{0,1,\cdots,q\},w \in \{0,\frac{1}{\tau},\cdots,\frac{e_b}{\tau}\}} \bigg\{ q
  +\beta\big[\rho \frac{\sigma^2}{h}(e^{\theta r}-1)+\Delta(r)-w\big]^{+} + \alpha
  \nonumber\\
  \times \mathbb{E}_{h,a,e}\left[V_{\alpha}(q-r+A,H,A,(e_b-w\tau+E)^{-},E)\right]
     \bigg\}.
  \end{eqnarray}
  We can see that
  $ V_{\alpha}(q,h,a,e_b,e)$ is increasing in $q$ and non-increasing in $e_b$ given $(h,a,e)$ since the larger the initial buffer the larger
  will be the cost to go, and the larger the initial battery energy the smaller will be the cost.\footnote{See the formal proof at Property
  \ref{Increasing in q} and Property \ref{Non-increasing in eb} in Appendix \ref{Properties of Valpha}.}
  Thus, $\arg \inf_{y \in \mathcal{X}}V_{\alpha}(y)=(0,h_0,a_0,E_{max},e_0):=x_0$, i.e., the infimum
  is obtained when the system begins with an empty buffer, a full battery, and for some channel sate $h_0$,
  arrival state $a_0$, and harvested energy arrival state $e_0$. When the buffer is empty, the set of feasible rate is $\{0\}$. Then $f(x_0,0,w)=0$,
  we get
\begin{eqnarray}
  V_{\alpha}(x_0)&=&\min_{w \in \{0,\frac{1}{\tau},\cdots,\frac{E_{max}}{\tau}\}}
 \alpha \mathbb{E}_{h_0,a_0,e_0}\left[V_{\alpha}(A,H,A,(E_{max}-w\tau+E)^-,E)\right] \nonumber\\
  &=& \alpha \mathbb{E}_{h_0,a_0,e_0}\left[V_{\alpha}(A,H,A,E_{max},E)\right].
\end{eqnarray}
Meanwhile, since policy $(q,0)$ is feasible for state $(q,h,a,e_b,e)$, then
\begin{eqnarray}\label{V(x) upper bound}
V_{\alpha}(x)\le q+\rho \frac{\sigma^2}{h}(e^{\theta q}-1)+C
+\alpha \mathbb{E}_{h,a,e}\left[V_{\alpha}(A,H,A,(e_b+E)^-,E)\right].
\end{eqnarray}
Let the system start in state $(a,h,a,e_b+e,e)$, we take the action $r[n]=a[n]$ and $w[n]<e[n]$ for all $n$.
Let $\xi(h,a,e_b,e)$ be the expected number of slots to hit the state $(a_0,h_0,a_0,E_{max},e_0)$.\footnote{When $w[n]<e[n]$, $E_{max}$ is the
absorbing state of the battery energy.} Observe that $\xi(h,a,e_b,e)$ is finite. Let
$
c_{max}=\max\limits_{h,a}\left\{ a+\rho \frac{\sigma^2}{h}(e^{\theta a}-1)\right\}+C.
$
Applying the Wald's lemma \cite{AMS45:A. Wald}, we get
 \begin{eqnarray}\label{Wald lemma}
  \lefteqn{
\alpha \mathbb{E}_{h,a,e}\left[V_{\alpha}(A,H,A,(e_b+E)^-,E)\right]\le c_{max}\xi(h,a,e_b,e)+ \alpha
}\nonumber\\
&\times&  \mathbb{E}_{h_0,a_0,e_0}\left[V_{\alpha}(A,H,A,E_{max},E)\right]
= c_{max}\xi(h,a,e_b,e)+V_{\alpha}(x_0).
 \end{eqnarray}
In (\ref{Wald lemma}), we have used $(E_{max}+E)^-=E_{max}$. Next,
combining (\ref{V(x) upper bound}) and (\ref{Wald lemma}), we have
$
V_{\alpha}(x)\le q+\rho \frac{\sigma^2}{h}(e^{\theta q}-1)+C
+ c_{max}\xi(h,a,e_b,e)+V_{\alpha}(x_0).
$
Thus,
$
V_{\alpha}(x)-V_{\alpha}(x_0)\le q+\rho \frac{\sigma^2}{h}(e^{\theta q}-1)+C
+ c_{max}\xi(h,a,e_b,e) < \infty.
$
Third, there exits a policy $\pi \in \mathcal{A}$ and an initial state $x \in \mathcal{X}$ such that $J_{x}^{\pi}(\beta)<\infty$ in the practical
problem. Otherwise, the cost is infinite for all policies and any policy is optimal.
Based on the above analysis, the conditions in Theorem 3.8 in \cite{MOR93:M. Schal} hold, and then we prove the lemma.

\subsection{Optimal policy for the discount cost MDP}

\subsubsection{Properties of $V_{\alpha}(q,h,a,e_b,e)$}\label{Properties of Valpha}
Property \ref{Increasing in q} - Property \ref{convexity} give the properties of the value function $V_{\alpha}(q,h,a,e_b,e)$.
\begin{property}\label{Increasing in q}
$V_{\alpha}(q,h,a,e_b,e)$ is an increasing
function of $q$.
\end{property}
\begin{IEEEproof}
We verify the increasing property by induction.
The value iteration algorithm (or successive approximation method) corresponding to (\ref{DCOE ueta}) is
 \begin{eqnarray} \label{iteration for V_alpha}
 V_{\alpha,n}(q,h,a,e_b,e)&=&\min_{u\in \{0,1,\cdots,q\},\eta \in \{0,1,\cdots,e_b\}} \bigg\{ q
 +\beta\big[\rho \frac{\sigma^2}{h}(e^{\theta (q-u)}-1)+\Delta(q-u)-\frac{e_b-\eta}{\tau}\big]^+ +\alpha \nonumber\\
 &\times&  \mathbb{E}_{h,a,e}\left[V_{\alpha,n-1}(u+A,H,A,(\eta+E)^-,E)\right]
    \bigg\}
 \end{eqnarray}
 with $V_{\alpha,0}(q,h,a,e_b,e)=0.$
Accordingly, $V_{\alpha,0}=0$, and $V_{\alpha,1}=q$. The increasing property in $q$ holds. Assume $V_{\alpha,n-1}(q,h,a,e_b,e)$ is increasing in $q$.
Fix $(h,a,e_b,e)$, in the state $(q+1,h,a,e_b,e)$,
 the set of feasible $u$ is $\{0,1,\cdots,q+1\}$ whereas it is $\{0,1,\cdots,q\}$ for state $(q,h,a,e_b,e)$.
Consider state $(q+1,h,a,e_b,e)$, let the optimal action be $(u^*,\eta^*)$ with $u^* \in \{0,1,\cdots,q\}$, hence
$
V_{\alpha,n}(q+1,h,a,e_b,e)  = q +1+\beta
\big[\rho \frac{\sigma^2}{h}(e^{\theta (q+1-u^*)}-1)+\Delta(q+1-u^*)-\frac{e_b-\eta^*}{\tau}\big]^+
+ \alpha \mathbb{E}_{h,a,e}\left[V_{\alpha,n-1}(u^*+A,H,A,(\eta^*+E)^-,E)\right].
$
 As $(u^*,\eta^*)$ is feasible in state $(q,h,a,e_b,e)$,
 $
 V_{\alpha,n}(q,h,a,e_b,e) \le  q
 +\beta\big[\rho \frac{\sigma^2}{h}(e^{\theta (q-u^*)}-1)+\Delta(q-u^*)-\frac{e_b-\eta^*}{\tau}\big]^+
 + \alpha
 \mathbb{E}_{h,a,e}
  $
 $
 \left[V_{\alpha,n-1}(u^*+A,H,A,(\eta^*+E)^-,E)\right]
 \le V_{\alpha,n}(q+1,h,a,e_b,e).
$
 If $(u^*,\eta^*)$ with $u^* =q+1$,
 \begin{eqnarray}
V_{\alpha,n}(q+1,h,a,e_b,e)  =q +1
+ \alpha \mathbb{E}_{h,a,e}\left[V_{\alpha,n-1}(q+1+A,H,A,(\eta^*+E)^-,E)\right].
 \end{eqnarray}
Meanwhile, since $(q,\eta^*)$ is feasible in state $(q,h,a,e_b,e)$,
\begin{eqnarray}
 V_{\alpha,n}(q,h,a,e_b,e) &\le&   q
 + \alpha \mathbb{E}_{h,a,e}\left[V_{\alpha,n-1}(q+A,H,A,(\eta^*+E)^-,E)\right]
  \stackrel{(a)}{\le}V_{\alpha,n}(q+1,h,a,e_b,e),\nonumber
 \end{eqnarray}
 where (a) holds since the induction hypothesis.
\end{IEEEproof}
\begin{property}\label{Non-increasing in eb}
$V_{\alpha}(q,h,a,e_b,e)$ is a non-increasing
function of $e_b$.
\end{property}
\begin{IEEEproof}
We verify this by induction. According to (\ref{iteration for V_alpha}), $V_{\alpha,0}=0$, and then $V_{\alpha,1}=q$. The non-increasing property
holds. Assume $V_{\alpha,n-1}(q,h,a,e_b,e)$ is non-increasing in $e_b$.
Given $(q,h,a,e)$, consider state $(q,h,a,e_b,e)$, let $(u^*,\eta^*)$ be the optimal policy, i.e.,
$
V_{\alpha,n}(q,h,a,e_b,e)  =q
+\beta\big[\rho \frac{\sigma^2}{h}(e^{\theta (q-u^*)}-1)+\Delta(q-u^*)-(e_b-\eta^*)/\tau\big]^+
+\alpha \mathbb{E}_{h,a,e}\left[V_{\alpha,n-1}(u^*+A,H,A,(\eta^*+E)^-,E)\right].
$
For state $(q,h,a,e_b+1,e)$,  $(u^*,\eta^*)$ is feasible, then we have
$
V_{\alpha,n}(q,h,a,e_b+1,e) \le   q+\beta\big[\rho
 \frac{\sigma^2}{h}(e^{\theta (q-u^*)}-1)+\Delta(q-u^*)-(e_b+1-\eta^*)/\tau\big]^+
+ \alpha \mathbb{E}_{h,a,e}\left[V_{\alpha,n-1}(u^*+A,H,A,(\eta^*+E)^-,E)\right]
\le V_{\alpha,n}(q,h,a,e_b,e).
$
\end{IEEEproof}
\par
In the practical case, the allocated harvested power will not surpass the required total power. Thus, we assume the $(u,\eta)$ always guarantees that
\begin{eqnarray}\label{Assumption1}
\rho \frac{\sigma^2}{h}(e^{\theta (q-u)}-1)+\Delta(q-u)\ge\frac{e_b-\eta}{\tau}.
\end{eqnarray}
Based on this assumption,
$P_{grid}(x,r,w)=P(x,r)-w$. The following property gives the joint convexity of $V_{\alpha}(q,h,a,e_b,e)$ in $(q,e_b)$.
\begin{property}\label{convexity}
$V_{\alpha}(q,h,a,e_b,e)$ is convex in $(q,e_b)$.
\end{property}
\begin{IEEEproof}
First, we prove the following claim.
\begin{claim}\label{proinappendix}
For $\phi \in (0,1)$ and $\forall x_1,x_2,y$, $\phi\min\{x_1,y\}+(1-\phi)\min\{x_2,y\}\le \min\{\phi x_1+(1-\phi)x_2,y\}$.
\end{claim}
\begin{IEEEproof}
The claim can be proved by considering $\min\{x_1,x_2\}>y$, $\max\{x_1,x_2\}<y$, and $\min\{x_1,x_2\}\le y\le \max\{x_1,x_2\}$, respectively.
\end{IEEEproof}
The convexity is proved by induction.
For $n=0$, $V_{\alpha,0}=0$, and it is convex. Assume $ V_{\alpha,n-1}(q,h,a,e_b,e)$ is convex in $(q,e_b)$. Fix $(q,h,a,e_b,e)$, let $(u_1,\eta_1)$
and $(u_2,\eta_2)$ be the optimal policy for $(q_1,e_{b1})$ and $(q_2,e_{b2})$. Then, we get
\begin{eqnarray}
  \lefteqn{
\phi V_{\alpha,n}( q_1,h,a, e_{b1},e)+(1-\phi)V_{\alpha,n}(q_2,h,a,e_{b2},e)= \phi\Big[ q_1+\beta(\rho \frac{\sigma^2}{h}(e^{\theta ( q_1-u_1)}-1)
}
  \nonumber\\
  &+&\Delta(q_1-u_1)
-\frac{e_{b1}-\eta_1}{\tau})\Big] +
(1-\phi)[q_2+\beta(\rho \frac{\sigma^2}{h}(e^{\theta (q_2-u_2)}-1)
+\Delta( q_2-u_2)-\frac{e_{b2}-\eta_2}{\tau})] \nonumber\\
&+& \alpha \mathbb{E}_{h,a,e}\Big[\phi V_{\alpha,n-1}(u_1+A,H,A,(\eta_1+E)^-,E)
+(1-\phi)V_{\alpha,n-1}(u_2+A,H,A,(\eta_2+E)^-,E)\Big] \nonumber\\
&\stackrel{(b)}{\ge}& \phi q_1+(1-\phi)q_2
+\beta\bigg[\rho \frac{\sigma^2}{h}(e^{\theta [\phi( q_1-u_1)+(1-\phi)( q_2-u_2)]}-1)
+\Delta(\phi( q_1-u_1)+(1-\phi)( q_2-u_2))\nonumber\\
&-&\frac{1}{\tau}(\phi( e_{b1}-\eta_1)+(1-\phi)( e_{b2}-\eta_2))\bigg]
+\alpha \mathbb{E}_{h,a,e}\bigg[ V_{\alpha,n-1}(\phi u_1+(1-\phi)u_2
+A,H,A,\phi (\eta_1+E)^-\nonumber\\
&+&(1-\phi)(\eta_2+E)^-,E)
\stackrel{(c)}{\ge} \phi q_1+(1-\phi)q_2
+\beta\bigg[\rho \frac{\sigma^2}{h}(e^{\theta [\phi( q_1-u_1)+(1-\phi)( q_2-u_2)]}-1)+\Delta(\phi( q_1-u_1)\nonumber\\
&+&(1-\phi)( q_2-u_2))-\frac{1}{\tau}(\phi( e_{b1}-\eta_1)+(1-\phi)( e_{b2}-\eta_2))\bigg]
+\alpha \mathbb{E}_{h,a,e}\bigg[ V_{\alpha,n-1}(\phi u_1+(1-\phi)u_2 \nonumber
\end{eqnarray}
\begin{eqnarray}
&+&A,H,A,(\phi \eta_1+(1-\phi)\eta_2+E)^-,E)
\stackrel{(d)}{\ge}
V_{\alpha,n}( \phi q_1+(1-\phi)q_2,h,a, \phi e_{b1}+(1-\phi) e_{b2},e),\nonumber
\end{eqnarray}
where (b) holds because of the convexity of $e^{\theta (q-u)}+\Delta(q-u)$ (with respect to $u$) and $ V_{\alpha,n-1}(q,h,a,e_b,e)$, (c) holds because
of Claim \ref{proinappendix} as well as Property \ref{Non-increasing in eb}, and
(d) holds since $(\phi u_1+(1-\phi)u_2,\phi \eta_1+(1-\phi)\eta_2)$ is feasible for $\phi( q_1,h,a, e_{b1},e)+(1-\phi)(q_2,h,a,e_{b2},e)$. The proof
completes.
\end{IEEEproof}

\subsubsection{On the discount optimal policy}\label{discount optimal}

For a state-action pair $\left(x=(q,h,a,e_b,e),(r,w)\right)\in \mathcal{X}\times \mathcal{A}(x)$,
define $u:=q-r$ and $\eta:=e_b-w\tau,$ i.e., let $u$ and $\eta$ denote the remaining data in the buffer and the remaining energy in the battery,
respectively. Then $(u(x),\eta(x))$ can also define a stationary policy. We can analysis the policy in terms of the remaining data in the buffer $u$
and the remaining energy in the battery $\eta$.
\begin{proposition}\label{Necessary condition for the discout optimality}
Denote the discount optimal policy in state $x=(q,h,a,e_b,e)$ as $(u^*(x),\eta^*(x))$. Then, $(u^*(x),\eta^*(x))$
satisfies the following inequality array
\begin{eqnarray}\label{u}
Z_1(q,u^*,h,a,\eta^*,e) \le \beta\rho \frac{\sigma^2}{h}e^{\theta q}(e^{\theta}-1)
 \le Z_1(u^*+1,h,a,\eta^*,e),
\end{eqnarray}
\begin{eqnarray}\label{eta}
Z_2(u^*,h,a,\eta^*,e)\le \frac{-\beta}{\tau}
\le Z_2(u^*,h,a,\eta^*+1,e),
\end{eqnarray}
\begin{eqnarray}\label{u eta}
Z_3(q,u^*,h,a,\eta^*,e) \le \beta\rho \frac{\sigma^2}{h}e^{\theta q}(e^{\theta}-1)
 \le Z_3(u^*+1,h,a,\eta^*+1,e),
\end{eqnarray}
where
$
Z_1(q,u,h,a,\eta,e)
=e^{\theta u}\bigg[\alpha \mathbb{E}_{h,a,e}\big[G_{1}(u+A,H,A,(\eta+E)^-,E)\big]
+\beta\big[\Delta(q-u)-\Delta(q-u+1)\big]
\bigg]
$
with
$
G_1(q,h,a,e_b,e)=
V_{\alpha}(q,h,a,e_b,e)-V_{\alpha}(q-1,h,a,e_b,e)
$
being the partial backward difference of $V_{\alpha}$ regarding $q$.
$
Z_2(u,h,a,\eta,e)=
\alpha \mathbb{E}_{h,a,e}\big[G_2(u+A,H,A,(\eta+E)^-,E)\big]
$
with
$
G_2(q,h,a,e_b,e)=
V_{\alpha}(q,h,a,e_b,e)-V_{\alpha}(q,h,a,e_b-1,e)
$
being the partial backward difference of $V_{\alpha}$ regarding $e_b$.
$
Z_3(q,u,h,a,\eta,e)=
e^{\theta u}\bigg[\alpha \mathbb{E}_{h,a,e}\big[G_{12}(u+A,H,A,(\eta+E)^-,E)\big]
+\beta\big[\Delta(q-u)-\Delta(q-u+1)\big]+\frac{\beta}{\tau}\bigg]
$
with
$
G_{12}(q,h,a,e_b,e)=
V_{\alpha}(q,h,a,e_b,e)-V_{\alpha}(q-1,h,a,e_b-1,e)
$
being the backward difference of $V_{\alpha}$ regarding $(q, e_b)$.
\end{proposition}
\begin{IEEEproof}
First,
the discounted cost optimality equation becomes
\begin{eqnarray} \label{DCOE ueta}
V_{\alpha}(q,h,a,e_b,e)&=&\min_{u\in \{0,1,\cdots,q\},\eta \in \{0,1,\cdots,e_b\}} \bigg\{ q
+\beta\big[\rho \frac{\sigma^2}{h}(e^{\theta (q-u)}-1)+\Delta(q-u)-\frac{e_b-\eta}{\tau}\big]^{+}
\nonumber\\
&+& \alpha \mathbb{E}_{h,a,e}\left[V_{\alpha}(u+A,H,A,(\eta+E)^-,E)\right]
   \bigg\},
\end{eqnarray}
Let
$
S(u,\eta)=q
+\beta\Big[\rho \frac{\sigma^2}{h}(e^{\theta (q-u)}-1)+\Delta(q-u)-\frac{e_b-\eta}{\tau}\Big]
+ \alpha \mathbb{E}_{h,a,e}\left[V_{\alpha}(u+A,H,A,(\eta+E)^-,E)\right].
$
First, we have
\begin{eqnarray}
  \lefteqn{
S(u+1,\eta)-S(u,\eta)=\beta\rho \frac{\sigma^2}{h}(e^{\theta (q-u-1)}-e^{\theta (q-u)})+ \beta[\Delta(q-u-1)-\Delta(q-u)]
}
\nonumber\\
&+&\alpha \mathbb{E}_{h,a,e}\big[V_{\alpha}(u+1+A,H,A,(\eta+E)^-,E)
-V_{\alpha}(u+A,H,A,(\eta+E)^-,E)\big]
\end{eqnarray}
and
\begin{eqnarray}
  \lefteqn{
S(u-1,\eta)-S(u,\eta)=\beta\rho \frac{\sigma^2}{h}(e^{\theta (q-u+1)}-e^{\theta (q-u)})+ \beta[\Delta(q-u+1)-\Delta(q-u)]
}
\nonumber\\
&+&\alpha \mathbb{E}_{h,a,e}\big[V_{\alpha}(u-1+A,H,A,(\eta+E)^-,E)
-V_{\alpha}(u+A,H,A,(\eta+E)^-,E)\big].
\end{eqnarray}

Then applying $S(u^*+1,\eta^*)-S(u^*,\eta^*)\ge 0$ and $S(u^*-1,\eta^*)-S(u^*,\eta^*)\ge 0$, we obtain (\ref{u}).
Similarly, as
$
S(u,\eta+1)-S(u,\eta)=\frac{\beta}{\tau}+\alpha \mathbb{E}_{h,a,e}\big[V_{\alpha}(u+A,H,A,(\eta+1+E)^-,E)
-V_{\alpha}(u+A,H,A,(\eta+E)^-,E)\big]
$
and
$
S(u,\eta-1)-S(u,\eta)=\frac{-\beta}{\tau}
+\alpha \mathbb{E}_{h,a,e}\big[V_{\alpha}(u+A,H,A,(\eta-1+E)^-,E)
-V_{\alpha}(u+A,H,A,(\eta+E)^-,E)\big],
$
we can reach (\ref{eta}) from $S(u^*,\eta^*+1)-S(u^*,\eta^*)\ge 0$ and $S(u^*,\eta^*-1)-S(u^*,\eta^*)\ge 0$.
In addition,
\begin{eqnarray}
  \lefteqn{
S(u+1,\eta+1)-S(u,\eta)=\beta\rho \frac{\sigma^2}{h}(e^{\theta (q-u-1)}-e^{\theta (q-u)})+ \frac{\beta}{\tau}+\beta[\Delta(q-u-1)-\Delta(q-u)]
}
\nonumber\\
&+&\alpha\mathbb{E}_{h,a,e}\big[V_{\alpha}(u+1+A,H,A,(\eta+1+E)^-,E)
-V_{\alpha}(u+A,H,A,(\eta+E)^-,E)\big]
\end{eqnarray}
and
\begin{eqnarray}
\lefteqn{
S(u-1,\eta-1)-S(u,\eta)=\beta\rho \frac{\sigma^2}{h}(e^{\theta (q-u+1)}-e^{\theta (q-u)})
-\frac{\beta}{\tau}+\beta[\Delta(q-u+1)-\Delta(q-u)]
}
\nonumber\\
&+&\alpha \mathbb{E}_{h,a,e}\big[V_{\alpha}(u-1+A,H,A,(\eta-1+E)^-,E)
-V_{\alpha}(u+A,H,A,(\eta+E)^-,E)\big].
\end{eqnarray}
Then, (\ref{u eta}) can be obtained by applying $S(u^*-1,\eta^*-1)-S(u^*,\eta^*)\ge 0$ and $S(u^*+1,\eta^*+1)-S(u^*,\eta^*)\ge 0$.
\end{IEEEproof}
\emph{Remark:
When $(u^*,\eta^*)$ is on the boundary of the feasible set, corresponding conditions can also be obtained by following the proof of Proposition
\ref{Necessary condition for the discout optimality}.
}
\begin{proposition}\label{discount optimal for two special cases}
For $x=(q,h,a,e_b,e)$ satisfying
\begin{eqnarray} \label{1}
Z_1\big(q,0,h,a,\tau\max\{0,\frac{e_b}{\tau}-P(x,q)\},e\big)
> \beta\rho \frac{\sigma^2}{h}e^{\theta q}(e^{\theta}-1)
\end{eqnarray}
and
\begin{eqnarray}\label{2}
Z_2(0,h,a,\tau\max\{0,\frac{e_b}{\tau}-P(x,q)\},e)> \frac{-\beta}{\tau},
\end{eqnarray}
$(0,\tau\max\{0,\frac{e_b}{\tau}-P(x,q)\})$ is the discount optimal policy.
In addition, for $(q,h,a,e_b,e)$ satisfying
\begin{eqnarray}\label{3}
Z_1(q,q,h,a,e_b,e)< \beta\rho \frac{\sigma^2}{h}e^{\theta q}(e^{\theta}-1)
\end{eqnarray}
and
\begin{eqnarray}\label{4}
Z_2(q,h,a,e_b,e)< \frac{-\beta}{\tau},
\end{eqnarray}
$(q,e_b)$ is the discount optimal policy.
\end{proposition}
\begin{IEEEproof}
Using Property \ref{convexity} in Appendix \ref{Properties of Valpha}, we can derive that
$Z_1(q,u,h,a,\eta,e)\le Z_1(q,u+1,h,a,\eta,e)$, $Z_1(q,u,h,a,\eta,e)\le Z_1(q,u,h,a,\eta+1,e)$,
$Z_2(u,h,a,\eta,e)\le Z_2(u,h,a,\eta+1,e)$, $Z_2(u,h,a,\eta,e)\le Z_2(u+1,h,a,\eta,e)$, and  $Z_3(q,u,h,a,\eta,e)\le
Z_3(q,u+1,h,a,\eta+1,e)$.\footnote{It is assumed that $\alpha \mathbb{E}_{h,a,e}\big[G_{1}(q+A,H,A,(\eta+E)^-,E)\big]
-e^{-\theta}\alpha \mathbb{E}_{h,a,e}\big[G_{1}(q-1+A,H,A,(\eta+E)^-,E)\big]
\ge \beta C$ and $\alpha \mathbb{E}_{h,a,e}\big[G_{12}(q+A,H,A,(\eta+E)^-,E)\big]-e^{-\theta}\alpha
\mathbb{E}_{h,a,e}\big[G_{12}(q-1+A,H,A,(\eta+E)^-,E)\big]
+\frac{\beta}{\tau}(1-e^{-\theta})\ge \beta C $. This assumption can be definitely satisfied when $C$ is small.}
On the other hand, (\ref{Assumption1}) should be satisfied.
Thus, given $(q,h,a,e)$, $Z_1(q,0,h,a,\tau\max\{0,\frac{e_b}{\tau}-P(x,q)\},e)$,
$Z_2(0,h,a,$ $\tau\max\{0,\frac{e_b}{\tau}-P(x,q)\},e)$, and $Z_3(q,0,h,a,$
$\tau\max\{0,\frac{e_b}{\tau}-P(x,q)\},e)$ are the smallest respectively. Following the proof of proposition \ref{Necessary condition for the discout
optimality},
we can prove the first half of the proposition
by contradiction.
Specifically, suppose $(0,\tau\max\{0,\frac{e_b}{\tau}-P(x,q)\})$ is not the optimal solution, then $S(u^*-1,\eta^*)-S(u^*,\eta^*)\ge 0$ or
$S(u^*,\eta^*-1)-S(u^*,\eta^*)\ge 0$ should hold. We have
$
Z_1(q,0,h,a,\tau\max\{0,\frac{e_b}{\tau}-P(x,q)\},e)
<Z_1(q,u^*,h,a,\eta^*,e)
\le \beta\rho \frac{\sigma^2}{h}e^{\theta q}(e^{\theta}-1)
$
 or
 $
Z_2(0,h,a,\tau\max\{0,\frac{e_b}{\tau}-P(x,q)\},e)
<Z_2(u^*,h,a,\eta^*,e)
\le \frac{-\beta}{\tau},
$
and the contradiction occurs.
\par
We can prove the second half of the proposition similarly by using contradiction. First, given $(q,h,a,e)$, $Z_1(q,q,h,a,e_b,e)$ and
$Z_2(q,h,a,e_b,e)$ are the largest values of $Z_1$ and $Z_2$, respectively. Assume $(q,e_b)$ is not the optimal solution, then
$S(u^*+1,\eta^*)-S(u^*,\eta^*)\ge 0$ or $S(u^*,\eta^*+1)-S(u^*,\eta^*)\ge 0$ should be satisfied. Consequently, we get
$
Z_1(q,q,h,a,e_b,e) \ge
Z_1(q,u^*+1,h,a,\eta^*,e)
\ge\beta\rho \frac{\sigma^2}{h}e^{\theta q}(e^{\theta}-1)
$
or
$
Z_2(q,h,a,e_b,e)\ge Z_2(u,h,a,\eta^*+1,e)\ge \frac{-\beta}{\tau}.
$
The contradiction occurs then.
\end{IEEEproof}

\emph{Remark: In Proposition \ref{Necessary condition for the discout optimality} and Proposition \ref{discount optimal for two special cases}, to compute
$Z_i(\cdot)~ i=1,2,3$, we need to compute $V_{\alpha}(\cdot)$. It can be
  obtained by value iteration (\ref{iteration for V_alpha}). }

  \begin{proposition} \label{monot of discount optimal policy}
 Denote $x=(q,h,a,e_b,e)$. The discount optimal transmit rate policy $r(x)=q-u^*(x)$ is non-decreasing in $q$ and $e_b$, respectively; The discount optimal battery energy allocation policy
  $w(x)=e_b-\eta^*(x)$ is non-decreasing in $q$ and
 $e_b$, respectively.
 \end{proposition}
  \begin{IEEEproof}
  First, it is easy to see that $r(x)$ is nondecreasing in $e_b$ and $w(x)$ is non-decreasing in $q$. Next,
  we prove the non-decreasing of $r(x)$ in $q$ by contradiction.
  Consider two states $x_1=(q_1,h,a,e_b,e)$ and $x_2=(q_2,h,a,e_b,e)$. We write $r(x_1)$ and $r(x_2)$ as $r(q_1)$ and $r(q_2)$ for brevity.  Assume $q_1<q_2$ but $r(q_1)>r(q_2)$, then $0\le r(q_2)<r(q_1)\le q_1<q_2$. $r(q_2),w(q_2)$ is feasible in $x_1$ and $r(q_1),w(q_1)$ is feasible in $x_2$. Since $r(\cdot)$ and $w(\cdot)$ are optimal, we have
 \begin{eqnarray}\label{mono1}
 \lefteqn{
 q_1+\beta\big[\rho \frac{\sigma^2}{h}(e^{\theta r(q_1)}-1)+C-w(q_1)\big]
 }\nonumber\\
 &+& \alpha\mathbb{E}_{h,a,e}\left[V_{\alpha}(q_1-r(q_1)+A,H,A,(e_b-w(q_1)\tau+E)^{-},E)\right]
\nonumber\\
 &\le&
 q_1+\beta\big[\rho \frac{\sigma^2}{h}(e^{\theta r(q_2)}-1)+\Delta(r(q_2))-w(q_2)\big]  \nonumber\\
 &+& \alpha\mathbb{E}_{h,a,e}\left[V_{\alpha}(q_1-r(q_2)+A,H,A,(e_b-w(q_2)\tau+E)^{-},E)\right]
 \end{eqnarray}

   \begin{eqnarray}\label{mono2}
 \lefteqn{
 q_2+\beta\big[\rho \frac{\sigma^2}{h}(e^{\theta r(q_2)}-1)+\Delta(r(q_2))-w(q_2)\big]
 }\nonumber\\
 &+& \alpha\mathbb{E}_{h,a,e}\left[V_{\alpha}(q_2-r(q_2)+A,H,A,(e_b-w(q_2)\tau+E)^{-},E)\right]
 \nonumber\\
 &\le&
 q_2+\beta\big[\rho \frac{\sigma^2}{h}(e^{\theta r(q_1)}-1)+C-w(q_1)\big]\nonumber\\
  &+& \alpha\mathbb{E}_{h,a,e}\left[V_{\alpha}(q_2-r(q_1)+A,H,A,(e_b-w(q_1)\tau+E)^{-},E)\right]
 \end{eqnarray}
 Add (\ref{mono1}) and (\ref{mono2}), we have
  \begin{eqnarray}\label{mono3}
  \lefteqn{
  \mathbb{E}_{h,a,e}\left[V_{\alpha}(q_1-r(q_2)+A,H,A,(e_b-w(q_2)\tau+E)^{-},E)\right]
  }
  \nonumber\\
  &-&\mathbb{E}_{h,a,e}\left[V_{\alpha}(q_1-r(q_1)+A,H,A,(e_b-w(q_1)\tau+E)^{-},E)\right]
   \nonumber\\
  &>&
  \mathbb{E}_{h,a,e}\left[V_{\alpha}(q_2-r(q_2)+A,H,A,(e_b-w(q_2)\tau+E)^{-},E)\right]
  \nonumber\\
  &-&\mathbb{E}_{h,a,e}\left[V_{\alpha}(q_2-r(q_1)+A,H,A,(e_b-w(q_1)\tau+E)^{-},E)\right]
   \end{eqnarray}
  As $V_{\alpha}(q,h,a,e_b,e)$ is convex in $(q,e_b)$, $\mathbb{E}_{h,a,e}\left[V_{\alpha}(y+A,H,A,(z+E)^{-},E)\right]$ is convex in $(y,z)$.
 (\ref{mono3}) contradicts the convexity. Then we prove the non-decreasing of $r(x)$ in $q$. The non-decreasing of $w(x)$ in $e_b$ can be verified similarly.
  \end{IEEEproof}
\subsection{Proof of Lemma \ref{one-stepcost}}\label{Poof of one-stepcost}
The lemma can be proved intuitively as follows. Given a transmission rate, the required power is known from the inverse of (\ref{relation between
required power and rate}). Out of this power, as much as possible shall be supplied by the battery, since battery energy is \lq\lq free\rq\rq. In
other words, any policy that draws power from the grid while energy is still available in the battery cannot outperform an equivalent one which
strictly uses battery energy first, that has the same total power.
\subsection{Proof of Lemma \ref{optimalforgreedy}}\label{Poof of optimalforgreedy}
Since $r(q,h,a,e_b,e)$ is irrelevant to $e_b$, given rate policy $r(q,h,a)$, the rate is determined independent of the battery allocation in each
timeslot. Then greedy battery allocation is optimal for one-step cost in each timeslot according to lemma \ref{one-stepcost}. Thus, the greedy battery
allocation policy is the optimal for (\ref{UP_beta}).

\subsection{Proof of Lemma \ref{non-increasing in beta}}\label{Proof of lemma non-increasing in beta}
 Since the optimal policy of UP$_\beta$ is $g_\beta$, we have
 \begin{eqnarray}
 J^{g_{\beta}}(\beta+\lambda)- J^{g_{\beta}}(\beta) \ge J^{g_{\beta+\lambda}}(\beta+\lambda)- J^{g_{\beta}}(\beta) \ge
 J^{g_{\beta+\lambda}}(\beta+\lambda)- J^{g_{\beta+\lambda}}(\beta)
 \end{eqnarray}
 for any positive $\beta>0$ and $\lambda>0$. Thus,
   \begin{eqnarray}
\lambda K^{\beta} \ge J^{g_{\beta+\lambda}}(\beta+\lambda)- J^{g_{\beta}}(\beta) \ge  \lambda K^{\beta+\lambda}>0.
    \end{eqnarray}
    The monotonicity of $J^{g_\beta}(\beta)$ and  $K^{g_\beta}$ with respect to $\beta$ are verified. In the following, we prove the non-decreasing of $B^{g_\beta}$ in $\beta$. First,
   similarly as in \cite{TIT08: M. Goyal A. Kumar and V. Sharma}, we can prove that $u^*(x)$ is non-decreasing in $\beta$.
    Next, as $A[n]$ is an independent process, then using (\ref{evolution equation for the buffer length}), we claim that $B^{g_\beta}$ is also
    non-decreasing in $\beta$.

\subsection{Proof of Lemma \ref{lemma greedy optimality discussion}}\label{Proof of lemma greedy optimality discussion}
We can verify the lemma through (\ref{DCOE}) together with Lemma \ref{discout to UP}.
When $\beta \gg 1$, we have $\beta \gg \alpha$.
Then
$V_{\alpha}=\min\limits_{r\in \{0,1,\cdots,q\},w \in \{0,\frac{1}{\tau},\cdots,\frac{e_b}{\tau}\}}$
 $\Big\{ q
  +\beta\big[\rho \frac{\sigma^2}{h}(e^{\theta r}-1)+\Delta(r)-w\big]^{+}\Big\}$. Given rate $r(x)$, we have $w(x)=\min\{\frac{e_b}{\tau},P(x,r)\}$
  (i.e., greedy policy) is discount optimal for state $x$.
  When $\beta$ is sufficient small, we have $\beta \ll \alpha$. Thus, $V_{\alpha}=\min\limits_{r\in \{0,1,\cdots,q\},w \in
  \{0,\frac{1}{\tau},\cdots,\frac{e_b}{\tau}\}} \big\{q+\alpha\mathbb{E}_{h,a,e}\left[V_{\alpha}(q-r+A,H,A,(e_b-w\tau+E)^{-},E)\right]\big\}$. Using Property \ref{Non-increasing in eb} in Appendix \ref{Properties of Valpha}, $w=0$ is discount optimal.
Since limitation will not change the partial order, utilizing the second half of Lemma \ref{discout to UP}, we reach the lemma.
\subsection{Proof of Lemma \ref{dimension reduction under a condition}}\label{Proof of dimension reduction under a condition}
Since the constrained MDP (\ref{original MDP}) is equivalent to UP$_{\beta_0}$ or the constrained optimal policy is a mixed policy of optimal policies for UP$_{\beta^+}$ and UP$_{\beta^-}$. When $\beta_0 \gg 1$ or $\beta^-\gg 1$, according to the first half of Lemma \ref{lemma greedy optimality discussion}, we can derive the greedy policy is the optimal battery power allocation policy under given rate policy. Fix the greedy
policy as the battery power allocation policy in (\ref{original MDP}), we arrive at (\ref{original optimization problem converted to policy only r})
for solving the optimal rate policy.

%
%

\ifCLASSOPTIONcaptionsoff
  \newpage
\fi

\end{document}